\documentclass[useAMS, usenatbib]{mn2e}
\usepackage{amsmath}
\usepackage{graphicx}
\usepackage{epstopdf}

\def\be{\begin{equation}}
\def\ee{\end{equation}}
\def\ba{\begin{eqnarray}}
\def\ea{\end{eqnarray}}
\def\sl{\mathrm{sl}}

\def\max{\mathrm{max}}
\newcommand{\abar}{\bar{\mathcal{A}}}
\newcommand{\bbar}{\bar{\mathcal{B}}}
\newcommand{\pl}{\Omega_{\rm{pl}}}
\newcommand{\ps}{\Omega_{\rm{ps}}}
\newcommand{\plo}{ \Omega_{\rm{pl,0}}}

\newcommand{\thetaSL}{\theta_{\rm{sl}}}
\newcommand{\thetaSB}{\theta_{\rm{sb}}}
\newcommand{\thetaLB}{\theta_{\rm{lb}}}
\newcommand{\thetadLB}{\dot{\theta}_{\rm{lb}}}
\newcommand{\thetaLBo}{\theta_{\rm{lb,0}}}
\newcommand{\thetaSLo}{\theta_{\rm{sl,0}}}
\newcommand{\phiSL}{\phi_{\rm{sl}}}

\newcommand{\hatS}{\hat{{\bf S}}}
\newcommand{\hatL}{\hat{{\bf L}}}
\newcommand{\hatLb}{\hat{{\bf L}}_b}

\def\go{\mathrel{\raise.3ex\hbox{$>$}\mkern-14mu
             \lower0.6ex\hbox{$\sim$}}}

\def\lo{\mathrel{\raise.3ex\hbox{$<$}\mkern-14mu
             \lower0.6ex\hbox{$\sim$}}}

\begin{document}
\title[Dynamics of Stellar Spin]
{Dynamics of Stellar Spin Driven by Planets Undergoing 
Lidov-Kozai Migration: Paths to Spin-Orbit Misalignment}
\author[Natalia I. Storch, Dong Lai, and Kassandra R. Anderson]
{Natalia I. Storch$^{1,2}$\thanks{Email: natalia@tapir.caltech.edu},
Dong Lai$^2$ and Kassandra R. Anderson$^2$ \\
$^1$ TAPIR, Walter Burke Institute for Theoretical Physics, Mailcode 350-17, Caltech, Pasadena, CA 91125, USA\\
$^2$ Cornell Center for Astrophysics and Planetary Science, Department
of Astronomy, Cornell University, Ithaca, NY 14853, USA}


\label{firstpage}
\maketitle

\begin{abstract}
Many exoplanetary systems containing hot Jupiters (HJs) exhibit
significant misalignment between the spin axes of the host stars and
the orbital angular momentum axes of the planets (``spin-orbit
misalignment'').  High-eccentricity migration involving Lidov-Kozai
oscillations of the planet's orbit induced by a distant perturber is a
possible channel for producing such misaligned HJ systems.  Previous
works have shown that the dynamical evolution of the stellar spin axis during the
high-$e$ migration plays a dominant role in generating the observed
spin-orbit misalignment.  Numerical studies have also revealed various
patterns of the evolution of the stellar spin axis leading to the
final misalignment.
Here we develop an analytic theory to elucidate the evolution of
spin-orbit misalignment during the Lidov-Kozai migration of planets in
stellar binaries. Secular spin-orbit resonances play a key role in the 
misalignment evolution. We include the effects of short-range forces and
tidal dissipation, and categorize the different possible paths to
spin-orbit misalignment as a function of various physical parameters
(e.g. planet mass and stellar rotation period). We identify five
distinct spin-orbit evolution paths and outcomes, only two of which
are capable of producing retrograde orbits.  We show that these paths
to misalignment and the outcomes depend only on two dimensionless
parameters, which compare the stellar spin precession frequency
with the rate of change of the planet's orbital axis, and the
Lidov-Kozai oscillation frequency. Our analysis reveals a number of novel
phenomena for the stellar spin evolution, ranging from bifurcation, 
adiabatic advection, to fully chaotic evolution of spin-orbit angles.
\end{abstract}


\section{Introduction}

The discovery of the misalignment between the orbital axes of hot Jupiters
(HJs; giant planets with orbital periods of $\sim 3$~days)
and the spin axes of their host stars (e.g. Hebrard et al. 2008, 2010;
Narita et al. 2009; Triaud et al. 2010; Winn et al. 2009; Albrecht et
al. 2012; Winn \& Fabrycky 2015) continues to pose a significant puzzle.
While primordial disk misalignment (with respect to the stellar spin)
is a possible explanation (e.g. Bate et al. 2010; Lai et al.~2011;
Foucart \& Lai 2011; Batygin 2012; Batygin \& Adams 2013; Lai 2014;
Spalding \& Batygin 2014), it is likely that a significant fraction of
HJs and the associated spin-orbit misalignments are produced by
dynamical means involving multi-planet interactions or planet-binary interactions.

Lidov-Kozai (LK) oscillations (Lidov 1962, Kozai 1962) induced by
external stellar or planetary companions -- one of the proposed channels of hot Jupiter
formation (e.g.  Wu \& Murray 2003; Fabrycky \& Tremaine 2007; Correia
et al 2011; Beaug{\'e} \& Nesvorn{\'y} 2012; Naoz et al. 2012; 
Petrovich 2015; Anderson, Storch \& Lai 2016;
Mu{\~n}oz, Lai \& Liu 2016; Petrovich \& Tremaine 2016) --
provide a natural means of generating spin-orbit
misalignment. Lidov-Kozai oscillations occur when the proto-HJ's host
star has a binary (or external planetary) companion. 
A proto-HJ is assumed to form at several
AUs from its host. If its orbital axis is sufficiently misaligned
relative to the outer binary axis, its orbit undergoes large
correlated variations in eccentricity and inclination. If the
misalignment is substantial, very high eccentricities (in excess of
$0.95$) can be attained; during these high-eccentricity phases, tidal
dissipation at periastron brings the planet close to its host,
eventually creating a HJ.

Since the host star of a giant planet generally has an appreciable rotation
(with rotation period ranging from a few days to 30~days) and is oblate,
significant coupling can exist between the orbital dynamics of the
proto-HJ and the dynamics of the stellar spin axis. This coupling is
vital in determining the final spin-orbit misalignments of these
systems. Indeed, in Storch, Anderson \& Lai (2014; hereafter SAL14) we
showed that the evolution of the stellar spin axis, driven by the
quasi-periodic changes in the planet orbit, can be very complex and
even chaotic, and this evolution depends sensitively on the planet
mass and stellar rotation period. Subsequently, in Storch \& Lai
(2015; hereafter SL15), we studied the origin of the chaotic behavior
(in terms of secular spin-orbit resonances and their overlaps) by
analysing the non-dissipative (i.e. no tidal dissipation) ``stellar spin +
planet + binary'' system and considering the regime in which the
stellar spin precession rate was much higher than the LK oscillation
frequency (the ``adiabatic'' regime).  In Anderson, Storch \& Lai
(2016; hereafter ASL16), we conducted a comprehensive population synthesis study of HJ
formation via LK migration in stellar binaries, including all relevant
physical effects (the octupole potential from the binary companion,
various short-range forces, tidal dissipation, and stellar spin-down
due to magnetic braking). In particular, our extensive Monte-Carlo
experiments (see Section 4 of ASL16) revealed various
paths of spin-orbit evolution during LK migration.

In the present work, we develop an analytic theory to understand
the evolution of spin-orbit misalignment during LK migration
in stellar binaries. We extend the analysis of SL15 to the
non-adiabatic regime (i.e. the regime in which the star precesses
slowly). We account for various non-ideal effects
(such as periastron advance due to General Relativity and planet
oblateness) and include tidal dissipation in our ``stellar spin + planet +
binary'' system.  Our goal is to provide theoretical explanations for
the various paths to spin-orbit misalignment that LK oscillations can
induce and to shed light on how the final spin-orbit misalignments of
HJs are achieved during LK migration. Although our focus is on
stellar binary-induced LK migration, most of our results can be
adapted to the planet-induced LK migration scenario.

This paper is organized as follows. In Section 2, we briefly review
our previous work and introduce the most important concepts and
equations in LK-driven spin-orbit dynamics. In Section 3, we discuss
the effect that short-range forces have on the spin-orbit dynamics. In
Section 4, we examine the different regimes of non-dissipative spin
dynamics. In Sections 5 and 6, we include tidal dissipation and study
the various paths toward misalignment during LK migration. We identify
the key paremeters that determine different behaviors of the
misalignment evolution.  We discuss the limitations and uncertainties
of our work in Section 7 and summarize our key findings in Section 8.

Readers who are less interested in the technical details can go to 
Section 5.1 for a description of the five different spin-orbit
evolution paths (with more detailed explanations in Sections 5.2-5.6),
and Section 8 for a brief summary.

\section{Lidov-Kozai-Driven Spin Dynamics: Concepts and Equations}

\subsection{Lidov-Kozai oscillations}
\label{LKsub}

We consider a star of mass $M_\star$ hosting a planet of mass $M_p$
(such that $M_\star \gg M_p$), and a stellar binary companion with
mass $M_b$.  Note that, in all calculations presented in this paper,
we set $M_\star=M_b=1 M_\odot$.
 
Throughout this work we consider LK oscillations to quadrupole order only.
This is an important simplification/approximation in order to
facilitate our theoretical analysis.  Note, however, that our
extensive Monte-Carlo calculations (ASL16) have showed
that the dominant effect of the octupole potential of the binary
companion is to increase the tidal disruption efficiency of planets,
and that a majority of HJs are formed through the ``normal'' (quadrupole)
channel (see also Mu{\~n}oz et al.~2016; Petrovich 2015).
Thus, we assume that the host star and binary companion are in a fixed
circular orbit (which naturally leads to zero octupole potential) with
semi-major axis $a_b$, and the binary orbital axis $\hatLb$ defines
the invariant plane of the system.

The planet orbit is described by its semi-major axis $a$, eccentricity
$e$, and angular momentum vector ${\bf L}$, which is inclined relative
to $\hatLb$. We define $\cos \thetaLB \equiv \hatL \cdot \hatLb$. In
the LK mechanism, if the initial $\thetaLB^0$ satisfies
$40^\circ \lo \thetaLB^0 \lo 140^\circ$, the planet
orbit undergoes oscillations in $e$ and $\thetaLB$, as well as nodal
and periastron precessions, while conserving ${\bf L}\cdot\hatLb$. The
oscillations happen on a characteristic timescale $t_k$ given by
\be
t_k^{-1}={n}
\left(\frac{M_b}{M_\star}\right)\left(\frac{a}{a_b}\right)^3,
\ee 
where $n\equiv (G M_\star/a^3)^{1/2}$ is the mean motion frequency of
the planet. In the absence of short-range forces (see section 3), the
maximum eccentricity achieved during an LK cycle is given by
\be
e_{\rm max}=\left(1-{5\over 3}\cos^2\thetaLB^0\right)^{1/2}.
\label{emax}
\ee
Short-range forces tend to reduce $e_{\rm max}$ from this value, but do not 
change the characteristics of the LK oscillations.
An important quantity for our later analysis is
the frequency of eccentricity oscillations, which is given by
\be
n_e = \mathcal{K} t^{-1}_k,
\ee
where $\mathcal{K}$ is of order unity and depends on strengths of short-range forces
and the value of the initial eccentricity $e_0$
($2\pi/n_e$ is the LK eccentricity oscillation period). We also define the quantities
\be
\pl \equiv \plo f(e) \equiv \frac{d\Omega}{dt}, \quad \thetadLB \equiv \frac{d\thetaLB}{dt}
\ee 
as the nodal precession rate of $\hatL$ (with $\plo$ the nodal
precession rate at $e=0$), and the nutation rate of $\hatL$,
respectively. Each of these quantities is a strong function of eccentricity, and
therefore time; together, they serve as the ``driving forces'' for the stellar
spin dynamics. 

\subsection{Stellar Spin Precession}

Due to the star's rotation-induced quadrupole, the stellar spin axis
$\hatS$ experiences periodic torquing from the planet, which is
strongest at the maximum eccentricity points of the LK cycle. This
torque induces precession of $\hatS$ around the planet's orbital
angular momentum axis $\hatL$, governed by the equation
\be
\frac{d\hatS}{dt} = \ps \hatL\times\hatS,
\label{dsdt}
\ee
\noindent where
\be
\ps\equiv -\frac{3GM_p(I_3-I_1)}{2a^3(1-e^2)^{3/2}}\frac{\cos\thetaSL}{S}
\label{omps}
\ee
is the precession frequency. Here $I_3$ and $I_1$ are the principal
moments of inertia of the star, $S$ is the magnitude of the spin
angular momentum, and $\thetaSL$ is the angle between $\hatS$ and
$\hatL$. To separate out the $\thetaSL$ and $e$ dependencies, we
define a function $\alpha(t)$ via
\be
\ps(t)\equiv -\alpha(t)\cos\thetaSL = -\frac{\alpha_0}{\left[1-e(t)^2\right]^{3/2}}\cos\thetaSL,
\label{alpha}
\ee
\noindent where
\ba
\alpha_0
&=&{3GM_p(I_3-I_1)\over 2a^3 I_3\Omega_\star}\nonumber\\
&=&1.19\times10^{-8}\left({2\pi\over1\mathrm{yr}}\right)\left({2k_{q}\over k_\star}\right)\left({10^3M_p\over M_\star}\right)\left({\hat{\Omega}_\star\over0.05}\right)\times \nonumber \\
&&\times \left({a\over1\mathrm{AU}}\right)^{-3}\left({M_\star\over M_\odot}\right)^{1/2}\left({R_\star\over R_\odot}\right)^{3/2}.
\label{alpha0}
\ea
Here we have used $(I_3-I_1)\equiv k_q
M_\star R_\star^2\hat{\Omega}_\star^2$, 
with $\hat\Omega_\star=\Omega_\star/(GM_\star/R_\star^3)^{1/2}$ the dimensionless
stellar rotation rate, and $S=I_3\Omega_\star\equiv k_\star M_\star
R_\star^2\Omega_\star$. 
For a solar-type star, $k_q\approx 0.05$, and
$k_\star\approx 0.1$ (Claret \& Gimenez 1992). 

Note that since $\alpha(t)$ is just $\left|\ps\right|$ evaluated at
$\cos\thetaSL=1$, it gives the maximum attainable spin precession rate
during the LK cycle, and $\alpha_0$ gives the maximum precession rate
at $e=0$. We emphasize that these rates depends linearly on both the
planet mass $M_p$ and the stellar spin rate $\Omega_\star$, and are
strong inverse functions of the semi-major axis $a$.

SAL14 and SL15 further defined the ``adiabaticity parameter'' $\epsilon$ as
\be
\epsilon\equiv \frac{\plo}{\alpha_0},
\ee
and showed that, generically speaking, $\epsilon$ serves as a
predictor for the dynamical behavior of the system. Given a set of initial
parameters such that $\epsilon \gg 1$, the system behaves
``non-adiabatically'': the nodal precession of $\hatL$ (around
$\hatLb$) is much faster than the precession of $\hatS$ (around
$\hatL$); therefore, $\hatS$ essentially precesses around the time
average of $\hatL$, thereby conserving $\thetaSB$, the angle between
the stellar spin axis and the outer binary axis. If $\epsilon \ll 1$,
the system behaves ``adiabatically'': the nodal precession of $\hatL$
is much slower than the precession of $\hatS$, and therefore $\hatS$
has no trouble keeping up with $\hatL$ and $\thetaSL$ is
conserved. For intermediate values of $\epsilon$, which SAL14 termed the
``trans-adiabatic'' regime, the spin dynamics is complex and often
chaotic. SL15 focused on exploring the spin dynamics in this regime
but close to the adiabatic transition (i.e. for $\epsilon \lo 1$). We
summarize their methods and findings in the rest of this section.

\subsection{Hamiltonian Spin Dynamics}

Because our primary goal is to study the behavior of the spin-orbit
misalignment angle $\thetaSL$, it is convenient to work in a frame of
reference where $\hatL$ is invariant. In this frame, it can be shown
(SL15) that the spin dynamics are governed by the following Hamiltonian:
\ba
&&H(p,\phiSL,\tau)=\frac{\bar{\alpha}}{n_e}\biggl\{
-\frac{1}{2}p^2 +\epsilon \psi(\tau)\,p \nonumber\\
&&\qquad ~~-\epsilon \sqrt{1-p^2}\,\Bigl[\beta(\tau)\cos\phiSL
+\gamma(\tau)\sin\phiSL\Bigr]\biggr\}, 
\label{kozaiham}
\ea
where $\phi_{\rm sl}$ (the precession phase of $\hatS$ about $\hatL$)
and $p\equiv \cos\thetaSL$ constitute the conjugate
pair of variables (with $p$ acting as the conjugate momentum).
The rescaled time $\tau$ is defined as
\be
\tau(t) = \frac{n_e}{\bar{\alpha}} \int_0^t \alpha(t') dt', 
\ee
where
\be
\frac{\bar{\alpha}}{n_e}\equiv\frac{1}{2\pi}\int_0^{2\pi/n_e}\!\alpha(t)dt
\label{alphabar}
\ee
is the ratio of the time-averaged maximum spin precession frequency,
$\bar{\alpha}$, and the LK eccentricity oscillation frequency $n_e$
and thus gives the maximum number of times that $\hatS$ can go around
$\hatL$ in one LK cycle (see section 3). Thus, $\bar{\alpha}/n_e$ is,
in fact, a better measure of the adiabaticity of the system than
$\epsilon$; we discuss this in more detail later in the section. 
Note that $\tau$ is normalized such that
it varies from $0$ to $2\pi$ in one LK eccentricity cycle.
The dimensionless functions $\beta,\,\gamma\,\psi$ are given by
\ba
&&\epsilon\beta(\tau)=-\frac{\pl(\tau)}{\alpha(\tau)}\sin\thetaLB(\tau), \label{beta} \\
&&\epsilon\gamma(\tau)=\frac{\dot{\theta}_{\rm lb}(\tau)}{\alpha(\tau)}, \label{gamma}\\
&&\epsilon\psi(\tau)=-\frac{\pl(\tau)}{\alpha(\tau)}\cos\thetaLB(\tau) \label{psi}.
\ea
Since $\epsilon={\Omega_{\rm pl,0}/\alpha_0}$, the functions
$\beta(\tau)$, $\gamma(\tau)$ and $\psi(\tau)$ depend only on the
``shape'' of the orbit, i.e., on $e(\tau)$ (with $\tau$ varying from 0 to $2\pi$).  
These shape functions can then be decomposed into Fourier components, as
\ba
&&\beta(\tau)=\sum_{M=0}^{\infty} \beta_M \cos M\tau, \\
&&\gamma(\tau)=\sum_{M=1}^{\infty} \gamma_M \sin M\tau, \label{gamma2}\\
&&\psi(\tau) =\sum_{M=0}^{\infty} \psi_M \cos M\tau,
\ea
and the Hamiltonian can be written as 
\ba
&&H'=\frac{\bar{\alpha}}{n_e} \biggl\{-\frac{1}{2}p^2 
+ \epsilon\,\psi_0\, p + \epsilon\, p \sum_{M=1}^{\infty} \psi_M \cos Mt \nonumber \\
&&\qquad ~-\frac{\epsilon}{2}\sqrt{1-p^2}\sum_{M=0}^{\infty}\Bigl[
(\beta_M+\gamma_M)\cos(\phiSL- M\tau) \nonumber \\
&&\qquad\quad\quad\quad\quad+(\beta_M-\gamma_M)\cos(\phiSL+M\tau)\Bigr]\biggr\}. 
\label{hexp}
\ea
Note that $\gamma_0$ is not defined in Eq. (\ref{gamma2}). For
convenience, we set $\gamma_0$ = $\beta_0$, but note that in actuality
the time average of the function $\gamma(\tau)$ is $0$, due to its
antisymmetric shape.

A resonance occurs when the argument of one of the cosine functions in
the above Hamiltonian is slow-varying, i.e. if
\be
\frac{d\phiSL}{d\tau} = N,
\label{eq:reson}\ee
with $N$ an integer (positive or negative). All the discussion up to this
point applies for arbitrary $\epsilon$. 
We can appreciate the significance of (\ref{eq:reson}) by considering the small
$\epsilon$ limit.
In this case, the Hamiltonian (\ref{hexp}) is dominated by the
first term and we have $d\phiSL/d\tau \simeq - p \bar{\alpha}/n_e$.
Then the resonance condition becomes
\be
{\bar\Omega}_{\rm ps}
=-\bar{\alpha}\cos\thetaSL
\simeq  N n_e,\qquad {\rm with}~~N = 0,\pm 1,\pm 2,\pm 3,\cdots
\label{rescond}
\ee
That is, when the time-averaged stellar spin precession frequency 
equals an integer multiple of the LK eccentricity oscillation
frequency $n_e$, the system experiences a resonance. When this
happens, the influence of all other terms in Eq. (\ref{hexp}) can be
averaged out and the system is governed by the single-resonance
Hamiltonian
\ba
&& H_N = \frac{\bar{\alpha}}{n_e}\biggl[-\frac{1}{2}p^2 +\epsilon\,\psi_0 \,p\nonumber\\
&&\qquad - \frac{\epsilon}{2}\sqrt{1-p^2}\,
(\beta_N+\gamma_N)\cos(\phiSL-N\tau)\biggr].
\label{oneharmonic}
\ea
For a given $\bar{\alpha}$, a set of resonances are possible, with the zeroth-order resonant 
momenta given by 
\be
p_N=\left(\cos\theta_{\rm sl}\right)_N \simeq -\frac{N n_e}{\bar{\alpha}} = - \frac{N}{N_\max}.
\label{pr}
\ee
Since $\left|\cos\thetaSL\right|$ cannot {\it exceed} $1$, there
exists a maximum resonance order, $\lfloor \bar{\alpha}/n_e\rfloor$. We define
\be
N_{\rm max} = \frac{\bar{\alpha}}{n_e},
\label{nmax}
\ee
and allow it to be non-integer because, as discussed previously, it
also has physical significance as the maximum number of spin
precession cycles ($\hatS$ around $\hatL$) in one LK cycle. In later
sections we show that, when tidal dissipation is introduced, $N_\max$
is one of the two key parameters in determining the dynamical evolution of the
system.

As discussed in detail in SL15, in the $\cos\thetaSL$ vs $(\phiSL-N\tau)$
phase space, the region of influence of each resonance is defined by
its separatrix, which has a distinctive cat-eye shape centered on
$\cos\thetaSL=p_N$ and $\phiSL-N\tau=0$ or $\pi$, depending on the
sign of $(\beta_N+\gamma_N)$ (see Fig. \ref{shortrangeforces}). The
Chirikov criterion (Chirikov 1979) states that overlaps in the
separatrices of two or more resonances lead to chaos. We thoroughly
explored this idea in SL15 and showed that, indeed, the appearance of
chaos in the system can be explained by overlaps between resonances of
different $N$'s.

One final point needs to be made. Recall that the above discussion of
resonances applies in the regime where the first term of the
Hamiltonian (\ref{hexp}) dominates over the others.  This regime
corresonds roughly to $\epsilon \lo 1$.  More precisely, since in
general (see SL15) we have $(\beta_0 + \gamma_0) = 2 \beta_0 >
(\beta_N + \gamma_N)$ for $N \neq 0$, the boundary of this regime can
be defined more accurately as $2\epsilon\beta_0 \lo 1$. We thus define
\be
\bar{\mathcal{A}}\equiv\left(2 \epsilon \beta_0\right)^{-1}
\label{abar}
\ee
as a new, more precise, adiabaticity parameter, such that when $\abar
\go 1$ the system is adiabatic. Note that $\abar$ can be expressed as
\be
\abar = \frac{\bar\alpha}{2\langle \pl\sin\thetaLB\rangle} \equiv
\frac{N_\max}{2\hat{\Omega}_{\rm L}},
\label{abar2}
\ee
where the triangle brackets denote time averaging and we have defined
$\hat{\Omega}_{\rm L}\equiv \langle
\pl\sin\thetaLB\rangle/n_e$. Physically, 
$2\abar$ 
represents the ratio of the maximum average rate of spin precession to the average rate of
change in $\hatL$ during the LK cycle. Thus, if $\abar \ll 1$, the
spin vector hardly moves compared with the orbital angular momentum
vector and the system behaves non-adiabatically, whereas when $\abar
\gg 1$ the spin axis evolves adiabatically, closely following the
changing $\hatL$ (as long as $\cos\thetaSL$ is not very close to
$0$). Note that, because $\hat{\Omega}_{\rm L} \propto \cos\thetaLBo$,
for the high inclinations that we consider in this work it is in
general true that $\abar \go N_\max$.

The quantity $\hat{\Omega}_{\rm L}$ is invariant so long as the
``shape'' of the orbit remains the same, i.e. it depends only on
$e(\tau)$. Thus, so long as the ``shape'' of the orbit is unchanged,
only one parameter (either $\abar$ or $N_{\rm max}$) determines the
evolutionary behavior of the system. In comparing the behavior of
systems with orbits of different ``shapes'', however, both parameters
are necessary.

To recapitulate, $\abar$ measures the (LK-averaged) maximum spin precession
rate, $\bar{\alpha}$, relative to the rate of change of $\hatL$,
whereas $N_\max$ measures $\bar{\alpha}$ relative to the LK
eccentricity oscillation frequency $n_e$ and sets the maximum
resonance order. Both $N_\max$ and $\abar$ scale linearly with the
stellar spin rate and the planet mass (see Eq. \ref{alpha0}). In the
remainder of the paper, we will show how the values of these two
parameters determine the behavior of the system and, in the presence
of tidal dissipation, the ultimate fate of the spin-orbit misalignment
angle.

\section{Effect of Short-Range Forces}
\label{SRFsec}

The analysis of SL15 focused solely on the ``pure'' Lidov-Kozai system
with no short-range forces.
As a step toward realism, we
now account for extra periastron advances induced in the system by
various short-range forces, including GR and the tide- and
rotation-induced quadrupole moments of the planet (e.g. Wu \&
Murray 2003, Fabrycky \& Tremaine 2006, Liu et al. 2015). These extra
periastron advance terms affect the LK+spin dynamics in two ways.

First, they slightly change the LK eccentricity oscillation timescale
$n_e$; this change is small and has no effect on the system dynamics.

\begin{figure}
\centering
\scalebox{0.67}{\includegraphics{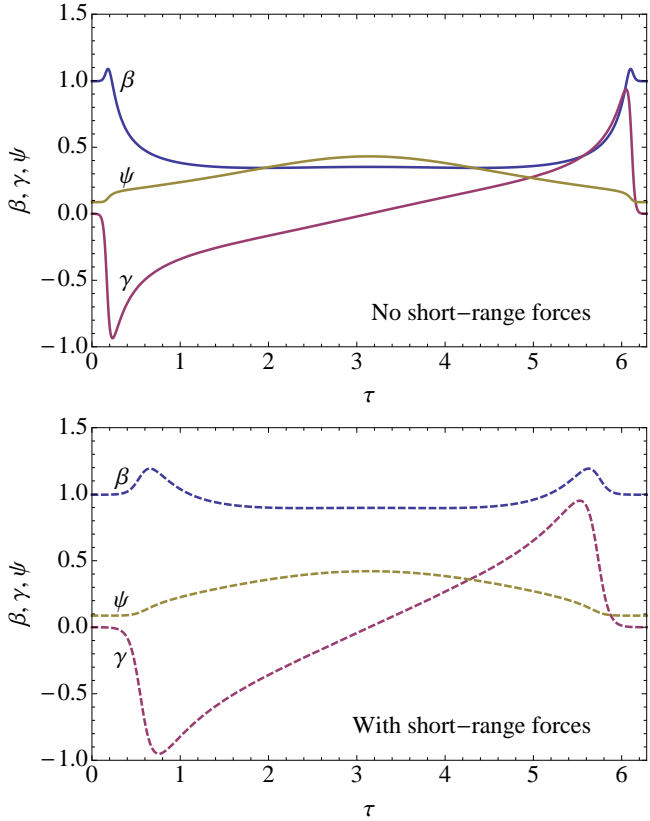}}
\caption{Shape functions $\beta$, $\gamma$, and $\psi$ for
  $\thetaLBo=85^\circ$ as a function of the rescaled time variable
  $\tau$, without (top panel) and with (bottom panel) short range
  forces. The physical parameters for the bottom panel are $M_p=5M_J$,
  $a_b=300$ AU, $a=1.5$ AU.}
\label{SRvsNoSR}
\end{figure}

Second, they reduce (sometimes significantly) the maximum eccentricity
attained during each LK cycle, thus changing the ``shape'' of
$e(t)$. In general, the new maximum eccentricity depends not only on
$\thetaLBo$ (cf. Eq. \ref{emax}), but also on the physical parameters
of the system, including the planet mass and radius ($M_p$ \& $R_p$),
the effective binary separation $a_b (1-e_b^2)^{1/2}$ (but recall that 
for the purposes of this paper we set $e_b=0$), and the planet semi-major axis $a$ (see Liu et al.~2015).
This leads to significant changes in the shape functions defined in
Eqs. (\ref{beta})-(\ref{psi}) and hence in the Fourier coefficients
$\beta_N$, $\gamma_N$ and $\psi_N$. Figure \ref{SRvsNoSR} demonstrates
this effect: In general, the variation in the shape functions becomes
smoother and less pronounced. We note that from Fig. \ref{SRvsNoSR} it
is obvious that short-range forces increase $\beta_0$ (the
$\tau$-average of Eq. \ref{beta}) and therefore decrease $\abar$
(Eq. \ref{abar}), making the system less adiabatic.

\begin{figure}
\centering
\scalebox{0.67}{\includegraphics{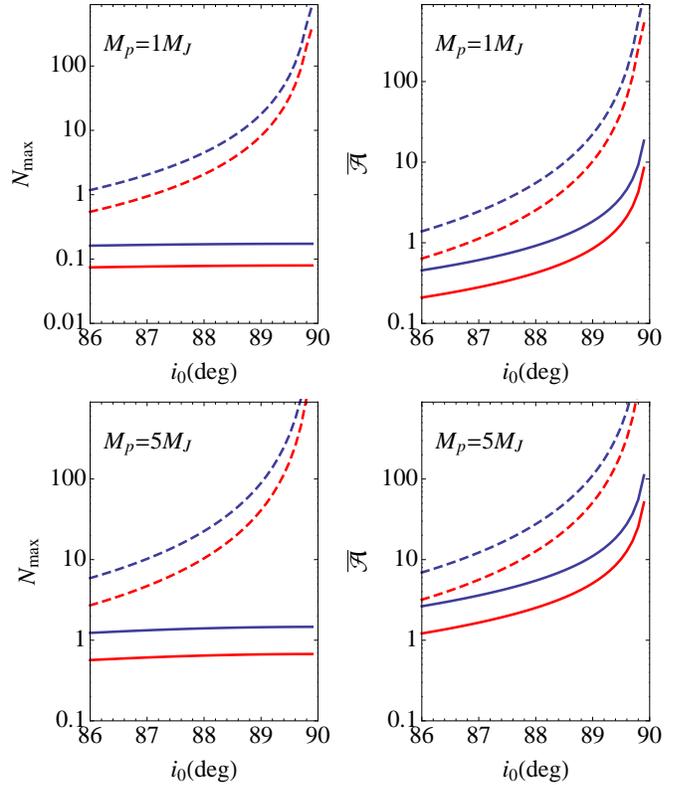}}
\caption{$N_{\rm max}$ (left panels) and $\abar$ (right panels) as a
  function of initial inclination for two planet masses (top panel:
  $M_p=1M_J$; bottom panel: $M_p=5M_J$) 
and two stellar spin periods
  (red: $P_\star=5$~days; blue: $P_\star=2.3$~days). Dashed: without
  short-range forces. Solid: with short-range forces.}
\label{Nmaxvsi}
\end{figure}

Due to the reduction of $e_\max$, the maximum of $\alpha(t)$
(Eq. \ref{alpha}) is also reduced, leading to a significant decrease
in $\bar{\alpha}$ (Eq. \ref{alphabar}). Thus, another consequence of
the inclusion of short-range forces is a decrease in the parameter
$N_\max$ (Eq. \ref{nmax}). Figure \ref{Nmaxvsi} (left) presents
$N_\max$ as a function of the initial orbital inclination $\thetaLBo$
with and without short-range forces. We see that, in general, $N_\max$
is greatly reduced when short-range forces are included. Furthermore,
because at high initial inclinations (e.g, $\thetaLBo \go 85$), the maximum
eccentricity is determined by the short-range forces and is not sensitive to
$\thetaLBo$ (e.g. Liu {\it et al.} 2015),
$N_\max$ becomes nearly independent of the initial inclination. It is
worth noting that $N_\max$ still scales linearly with the stellar spin
rate (or inversely with the spin period), but its dependence on $M_p$
is no longer as simple, since $M_p$ now plays a role in setting the
maximum eccentricity.

Likewise, the parameter $\abar$ is also affected (Fig. \ref{Nmaxvsi},
right). Like $N_\max$, it still scales linearly with the stellar spin
rate. However, unsurprisingly (cf. Eq. \ref{abar2}), it has a much
stronger dependence on the initial inclination.

\begin{figure}
\centering
\scalebox{0.55}{\includegraphics{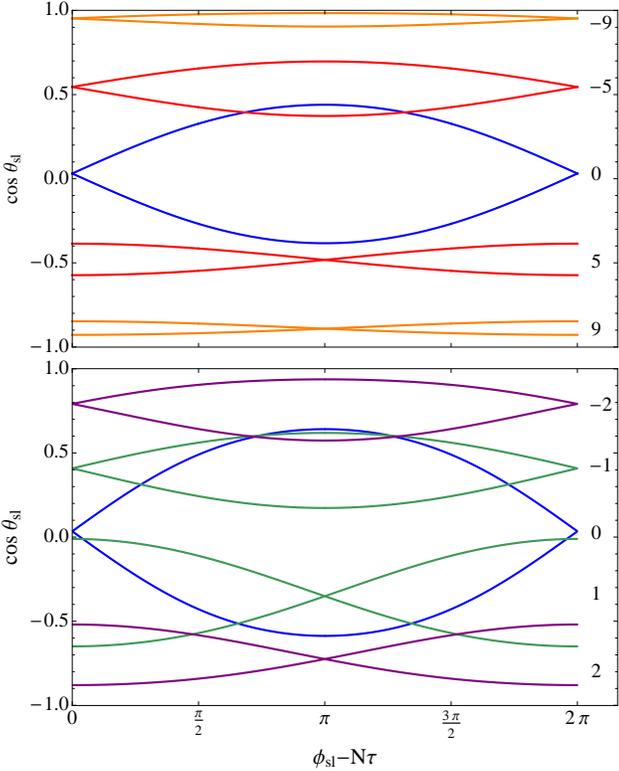}}
\caption{Separatrices in the phase space for a sample of resonances,
  labeled on the right-hand side with their corresponding $N$, for the
  shape functions presented in Fig. \ref{SRvsNoSR} (with
  $\thetaLBo=85^\circ$ and $\epsilon=0.1$), without (top panel) and
  with (bottom panel) short-range forces. Note that, for identical
  values of $\epsilon$ and $\thetaLBo$, inclusion of short-range
  forces significantly reduces the number of resonances affecting the
  system dynamics (i.e. the number of resonances that ``fit'' inside
  the phase space).}
\label{shortrangeforces}
\end{figure}

For comparison, Figure \ref{shortrangeforces} demonstrates how a
sample phase space previously presented in SL15 (Fig. 6) changes when
short-range forces are added to the system: As expected, the number of
resonances is significantly reduced, and on the whole most resonances
become wider.  In summary, given a system with a set of initial
parameters, the inclusion of short-range forces changes the shape
functions that drive the spin precession dynamics, and generally
decreases both $\abar$ and $N_\max$, reducing the degree of
adiabaticity of the system.

Finally, note that there is one more ``non-ideal'' effect we have
neglected: the perturbation of the planet's orbit due to the
rotation-induced stellar quadrupole. This perturbation comes in two
forms. First, the stellar quadrupole induces additional periastron
advance in the orbit, similar to the other short-range forces. 
Second, the planet's orbit experiences an extra nodal precession,
governed by the equation
\be
\left(\frac{d\hatL}{dt}\right)_{\rm SL} = \ps {S\over L}
\hatS\times\hatL,
\ee
where the ratio of the stellar spin angular momentum to the orbital
angular momentum is
\ba
&& {S\over L}\simeq 0.12 \left(\!{k_\star\over 0.1}\!\right)\!
\left(\!{M_\star\over M_\odot}\!\right)^{\!1/2}
\!\left(\!{R_\star\over R_\odot}\!\right)^2\left(\!{M_p\over M_J}\!\right)^{\!-1}
\nonumber\\
&&\qquad \times \left[{a(1-e^2)\over 0.05\,{\rm AU}}\right]^{\!-1/2}
\!\left(\!{P_\star\over 30\,{\rm d}}\!\right)^{\!-1}.
\label{eq:SoverL}\ea
We ignore these effects because, by creating feedback
between stellar spin precession and orbit precession, they break the
integrability of the Hamiltonian system by introducing more degrees of
freedom [i.e. $e(t)$, $\pl(t)$, etc. would no longer be solely
determined by LK dynamics and in general would not be ``known''
periodic forcing functions acting on the spin evolution].
Neglecting these feedback effects is the single biggest simplifying
assumption we make in our analysis. The significance of these effects
increases with increasing $S/L \propto \Omega_\star/M_p$ (see Section 4.3 of ASL16 for
further discussion). For low planet masses and high stellar rotation
rates, this makes our subsequent analysis of the spin dynamics
somewhat pedagogical. For higher planet masses and lower stellar spin
rates, however, the feedback is not as important, and our conclusions
should be fairly robust.

\section{Non-dissipative Regime Classification}
\label{nondiss}

In this section we examine the differences in the spin dynamical
behavior of a non-dissipative system in the non-adiabatic vs adiabatic
regimes.  We select two representative ``shapes'' of the LK orbit by
considering systems with $M_p=5 M_J$, $a=1.5$ AU, $a_b=300$ AU, and
either $\thetaLBo=89^\circ$ or $\thetaLBo=87^\circ$. We then vary the
stellar spin period, which scales up/down both $N_\max$ and
$\abar$. Note that, in order to explore the entire range of possible
behaviors, we consider a somewhat unphysical range of stellar spin
periods, from as large as $50$ days to as small as $1$ day.

We first consider the spin dynamics in the non-adiabatic regime, with $\abar \lo 1$
(section 4.1).
We then consider the adiabatic regime with $\abar \go 1$ (section 4.2), which is
further divided into two sub-regimes with $N_\max \lo 1$ and $N_\max
\go 1$ (recall that $\abar \go N_\max$). Finally, we specialize to the
dynamics of spin trajectories that start with $\cos\thetaSL=1$
(i.e. zero initial spin-orbit misalignment) and discuss their behavior
in each of the aforementioned regimes (section 4.3).

\begin{figure}
\centering
\scalebox{0.67}{\includegraphics{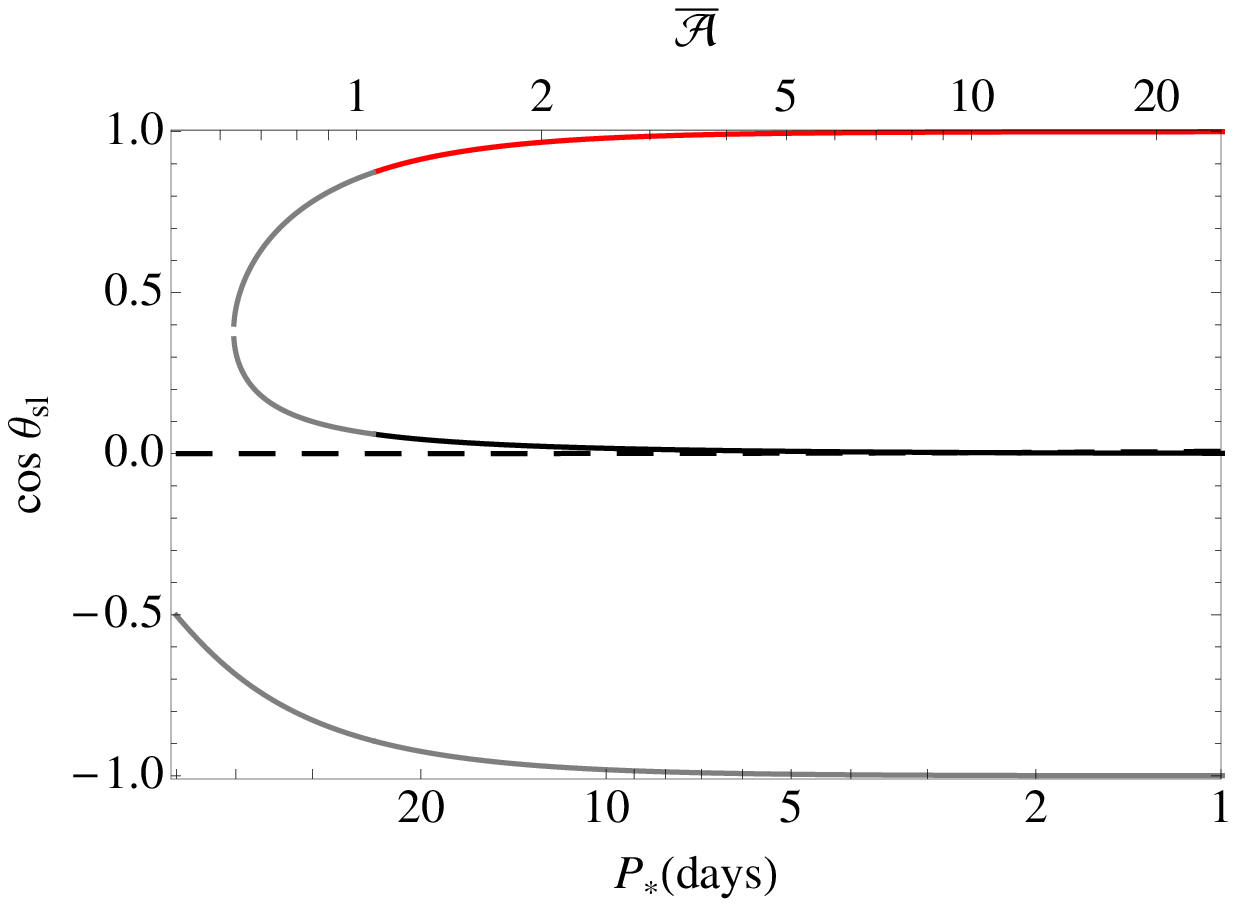}}
\caption{Locations of the fixed points of the Hamiltonian
  ($\ref{noham}$), for $\phiSL=0$ (solid lines) and $\phiSL=\pi$
  (dashed line), for $M_p=5 M_J$, $\thetaLBo=89^\circ$, as a function
  of the stellar spin period. For reference, the adiabaticity
  parameter $\abar$ is plotted on the top axis. $N_\max$ can be
  calculated as $N_\max \simeq \abar/7.63$. 
For more detail about the phase space structure, see
  Fig. \ref{samplephasespaces}. 
The colors of the lines match those of the corresponding separatrices in
Fig. \ref{samplephasespaces}.}
\label{fixedpoints}
\end{figure}

\subsection{Non-adiabatic Regime: $\abar \lo 1$}

\begin{figure*}
\centering
\includegraphics[width=\textwidth]{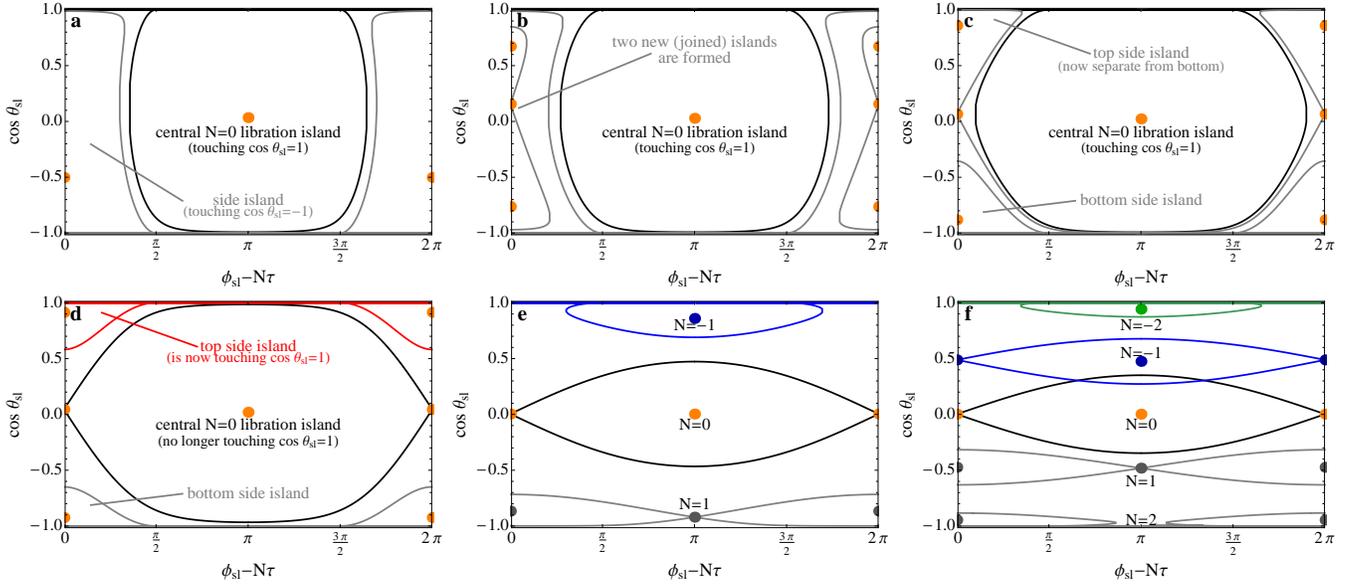}
\caption{Sample separatrices in phase space for $M_p=5 M_J$,
  $\thetaLBo=89^\circ$ and (from {\bf a} to {\bf f}), $P_\star = 50$,
  $35$, $25$, $20$, $3$, and $1.6$ days. The separatrices shown in
  gray are those that have no relevance to the dynamics of the
  initially-aligned trajectory ($\cos\thetaSLo=1$). The colored dots
  mark locations of fixed points. Here $\abar\simeq25.5/P_\star$ and
  $N_\max \simeq 3.34/P_\star$.}
\label{samplephasespaces}
\end{figure*}

The form of the Hamiltonian (\ref{hexp}) is such that the
non-adiabatic regime of spin dynamics does not easily lend itself to
pertubation theory and cannot be formally explored. Nevertheless, the
non-adiabatic regime is very important (especially for Jupiter-mass or
smaller planets). We therefore endeavor to study it based on physical
arguments we consider reasonable but not necessarily rigorous.

We classify the non-adiabatic regime as having $\abar \lo 1$. Since
typically $\hat{\Omega}_{\rm L} \lo 1$, this implies $N_\max \ll
1$. Thus in the non-adiabatic regime, the stellar spin vector $\hatS$
changes slowly compared to both the rate of change of $\hatL$ and the
LK oscillation frequency. We therefore surmise that most of the
relevant spin dynamics can be captured using a time-independent
Hamiltonian whose coefficients are the time averages of the externally
imposed shape functions ($\beta$, $\psi$, $\gamma$). In other words,
most of the spin dynamics can be understood by analyzing the $N=0$
Hamiltonian (cf. Eq.~\ref{oneharmonic})
\be
H_0 =N_\max\biggl[-\frac{1}{2}p^2 +\frac{p}{\bbar} - \frac{1}{2\abar}\sqrt{1-p^2}\,
\cos\phiSL\biggr],
\label{noham}
\ee
where we have replaced $\bar{\alpha}/n_e$ with $N_\max$, replaced
$\epsilon (\beta_0+\gamma_0)\equiv 2\epsilon\beta_0$ using the
definition of $\abar$, and have defined $\bbar \equiv (\epsilon\,\psi_0)^{-1}$.

First, it is useful to examine the fixed points of the Hamiltonian $H_0$,
and how they depend on the stellar spin period. By considering the
equations of motion, it is easy to see that this Hamiltonian has two
sets of fixed points: those with $\phiSL=0$ and those with
$\phiSL=\pi$. Figure \ref{fixedpoints} shows how the locations of
these fixed points in $\cos\thetaSL$ change with stellar spin
period. At very high spin periods, only two fixed points exist, one at
$\phiSL=0$ and another at $\phiSL=\pi$. These fixed points are closely
related to the well-known Cassini states (e.g. Fabrycky et
al.~2007). At $P_\star \lo 40$ days, another set of fixed points
appear for $\phiSL=0$. We can now examine how each of these fixed
points affects the system dynamics.

The top panels of Figure \ref{samplephasespaces} present example phase
spaces calculated based on the Hamiltonian (\ref{noham}), for which
$\abar \lo 1$. The curves shown in each panel are separatrices that
cannot be crossed by any spin evolution trajectory. 
The shapes of the separatrices constrain the possible trajectories. 
Thus, an analysis of the separatrix shapes sheds light on the possible
behaviors of the system.

At very low values of $\abar$ (Fig. \ref{samplephasespaces}, panel
{\bf a}), the phase space is roughly split into two islands of libration, each 
containing a fixed point.  The separatrix of the center island touches
$\cos\thetaSL=1$, whereas the other touches $\cos\thetaSL=-1$. A
trajectory starting inside one of these separatrices will librate
about the corresponding fixed point in the center of the
island. Trajectories starting in the narrow region in-between the two
separatrices are able to circulate.

As shown in Fig.~\ref{fixedpoints}, at $P_\star \lo 40$ days, two
more fixed points appear. Panel {\bf b} of
Fig.~\ref{samplephasespaces} shows the phase space
structure shortly after the appearance of these fixed points. Two new,
joined, libration islands appear. The bottom island has the previously
existing Cassini-state fixed point at its center. The top island has
one of the new fixed points at its center. The second new fixed point
defines the separatrix between these two islands.

As $\abar$ gets closer to $1$ (Fig.~\ref{samplephasespaces}, panel
{\bf c}), the central (centered on $\phiSL=\pi$) libration island
expands, and eventually cleaves the two side islands.

As $\abar$ further increases, the center libration island spans an
increasingly larger range of $\phiSL$, until, when $\abar \simeq 1$,
it spans the entire $\{0,2\pi\}$ range and detaches from
$\cos\thetaSL=1$, forming the standard cat-eye shaped $N=0$ resonance
(Fig.~\ref{samplephasespaces}, panel {\bf d}). (Note that the location
of the transition is {\it not} exact: as can be seen from
Fig.~\ref{samplephasespaces}, panel {\bf c}, the actual transition
happens slightly after $\abar \simeq 1$.) 
The fixed point that previously defined the separatrix for the top
side island now defines the separatrix for the cat-eye. At the same
time, the top side island {\it attaches} to $\cos\thetaSL=1$. This
marks the transition from non-adiabatic to adiabatic behavior.

\subsection{Adiabatic Regime: $\abar \go 1$}

\subsubsection{$N_\max \lo 1$}

Shortly after the non-adiabatic to adiabatic transition, $N_\max$ is
still small, and therefore the spin dynamics are still essentially
governed by the $N=0$ Hamiltonian. After the central $N=0$
island/resonance has detached from $\cos\thetaSL=1$, the top side
island merges upward and {\it attaches} to $\cos\thetaSL=1$
(Fig.~\ref{samplephasespaces}, panel {\bf d}, shown in red). As
$\abar$ and $N_\max$ continue to increase, this side island rapidly
shrinks and is soon overtaken in importance by the newly forming
$N=-1$ resonance. Likewise, the bottom island shrinks as well and is
soon dominated by the $N=1$ resonance.

\subsubsection{$N_\max \go 1$}

As $N_\max$ approaches $1$, the spin dynamics are no longer
determined solely by the $N=0$ Hamiltonian. Rather, the $N=1$ and $N=-1$ 
Hamiltonians must also be considered
(see Eq.~\ref{oneharmonic}). For $0.5 \lo N_\max \lo 1$, each of these
Hamiltonians produces a separatrix that is attached to
$\cos\thetaSL=1$ (for $N=-1$) and $\cos\thetaSL=-1$ (for $N=1$). As
$N_\max$ continues to increase, the separatrices ``emerge'' more fully
into the phase space until eventually they detach from the top and
bottom edges and form standard cat-eye shapes. Note that, due to the
slight asymmetry in the Hamiltonians introduced by the $p/\bbar$ term,
the bottom resonance detaches slightly earlier than $N_\max \simeq 1$,
whereas the top resonance detaches slightly later than $N_\max \simeq
1$ (Fig.~\ref{samplephasespaces}, panel {\bf e}).

After the $N=\pm 1$ resonances have emerged fully, the $N=\pm 2$
resonances begin to grow, and likewise detach and form cat-eye shapes
when $N_\max \simeq 2$ (Fig.~\ref{samplephasespaces}, panel {\bf f}).

\begin{figure}
\centering
\scalebox{0.67}{\includegraphics{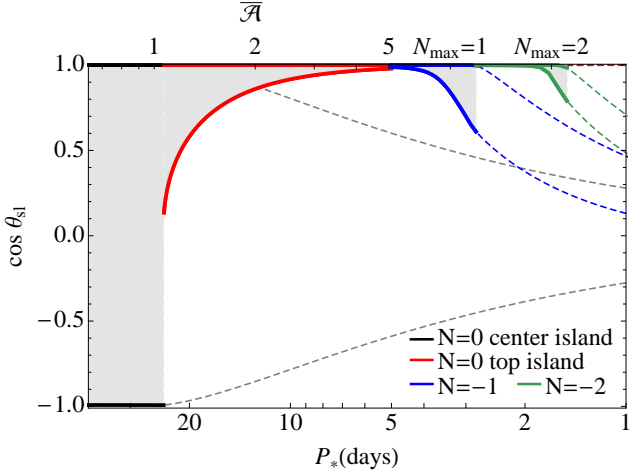}}
\caption{The range (in $\cos\thetaSL$) of each of the ``relevant''
  separatrices presented in Fig.~\ref{samplephasespaces} as a
  function of stellar spin period for $\thetaLBo=89^\circ$ and $M_p=5
  M_J$. Colors have been chosen to match those of
  Fig.~\ref{samplephasespaces}. The separatrix widths are shown in
  bold solid lines whenever they touch $\cos\thetaSL=1$ -- such separatrices
  then determine the behavior of an initially-aligned
  trajectory. After detaching from $\cos\thetaSL=1$, each separatrix
  is shown in thin dashed lines of the same color. Note that the transition
  from non-adiabatic behavior (controlled by the $N=0$ center island)
  to adiabatic behavior happens at $\bar{\mathcal{A}}\simeq 1$. In
  grey we show the maximum range of excursion that the
  initially-aligned trajectory can have. Also note that the transition from
  the red ($N=0$ top side island dominated) to blue ($N=-1$ dominated)
  is somewhat arbitrary, as there is no precise way to tell when the
  $N=-1$ resonance starts to dominate.}
\label{influence}
\end{figure}

\begin{figure}
\centering
\scalebox{0.67}{\includegraphics{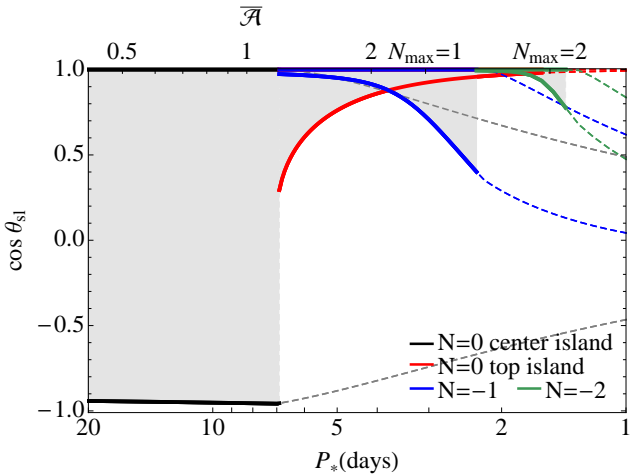}}
\caption{Same as Fig.~\ref{influence}, but with $\thetaSLo=87^\circ$.}
\label{influence87}
\end{figure}

\subsection{The initially-aligned trajectory}
\label{IATsub}

In the standard planetary system formation scenario, giant planets are
formed at a few AU's distance from their host stars, with the orbital
axis aligned with the stellar spin axis.
Although in recent years several methods of generating primordial
misalignment have been suggested (e.g. Bate et al. 2010; Batygin 2012;
Batygin \& Adams 2013; Lai 2014; Lai et al. 2011; Spalding \& Batygin
2014), for the remainder of this paper we will focus on systems with
$\cos\theta_{\rm sl,0} = 1$, i.e. no initial spin-orbit
misalignment. What determines the dynamic behavior of the stellar spin
as the planet undergoes LK oscillations?

One key observation can be taken away from the six panels presented in
Fig.~\ref{samplephasespaces}: regardless of what regime the system is
in, there is {\it always} a separatrix ``attached'' to
$\cos\thetaSL=1$. In the non-adiabatic regime ($\abar \lo 1$) this
separatrix is the $N=0$ center island. In the adiabatic regime ($\abar
\go 1$), this separatrix is the $N=0$ top side island 
when $N_\max\lo 0.5$ (note this number is somewhat arbitrary), 
and then $N=-1$ for $0.5\lo N_\max\lo 1$,
$N=-2$ for $1\lo N_\max\lo 2$, and so on.
These are the separatrices that determine the behavior of the
initially aligned trajectory.

Figures \ref{influence} and \ref{influence87} present a different way
of visualizing this information. They show the maximum vertical extent
of each of the relevant (``attached'' to $\cos\thetaSL=1$)
separatrices as a function of the stellar spin period $P_\star$, for
two different shape functions, with the corresponding values of
$\abar$ and $N_\max$ given on the top axes. Since, in each case, the
initially-aligned trajectory {\it starts out} on the relevant
separatrix, its maximum vertical width represents the maximum range of
spin-orbit misalignments that the trajectory can cover. Thus, in the
non-adiabatic regime the stellar spin has the most ``freedom'' and is
able to cover the largest range of $\cos\thetaSL$. As $\abar$
increases, the spin axis' range of excursion becomes progressively
more and more limited, though not monotonically so.

There is a clear difference between Figures \ref{influence}
($\thetaSLo=89^\circ$) and \ref{influence87} ($\thetaSLo=87^\circ$):
in Fig.~\ref{influence} the regions of influence of each separatrix
are clearly defined and well-separated. On the other hand, in
Fig.~\ref{influence87} it is not always clear which separatrix
dominates the evolution. Furthermore, Fig.~\ref{influence87} features
{\it overlaps} between the different separatrices, which, as we know,
is a signature of chaos (see SL15). 
We therefore expect that for $\thetaSLo=89^\circ$ the spin dynamics
for all stellar rotation periods should be regular and well-behaved,
whereas for $\thetaSLo=87^\circ$ chaotic behavior may arise for
certain ranges of rotation periods.

\begin{figure}
\centering
\scalebox{0.67}{\includegraphics{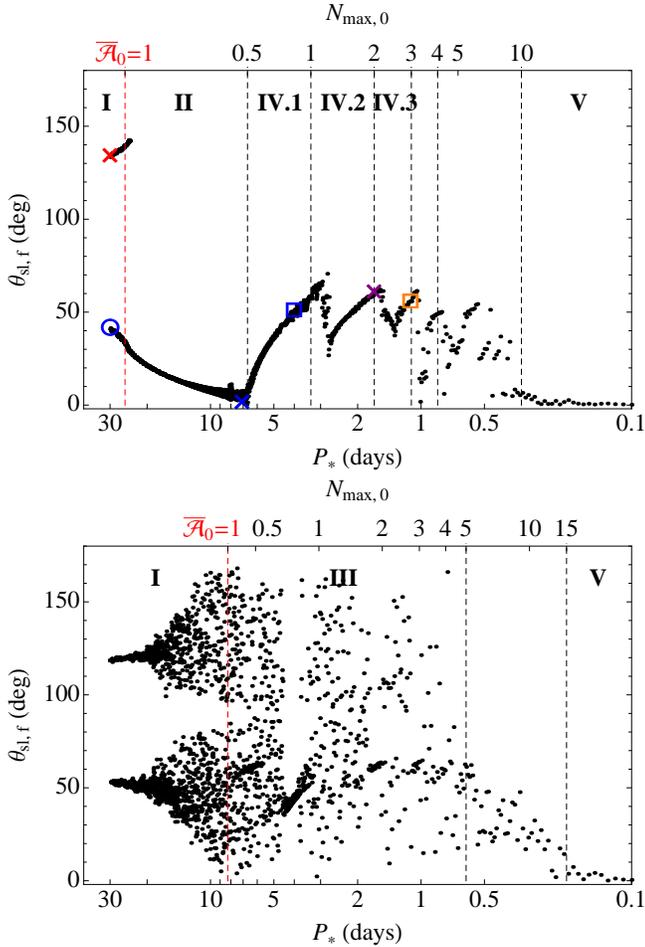}}
\caption{Final spin-orbit misalignment angle (after orbital decay and
  circularization due to tidal dissipation) as a function of the
  stellar spin period, for $M_p=5M_J$, $a_0=1.5$ AU, $a_b=300$ AU, and
  $\thetaLBo=89^\circ$ (top panel) or $\thetaLBo=87^\circ$ (bottom
  panel). The blue, red, and purple $\times$ symbols in the top panel
  correspond to the left, middle, and right example trajectories shown
  in Fig.~\ref{samplepaths}, respectively. The blue $\circ$ symbol
  corresponds to the blue trajectory in Fig. \ref{bimodality}. The
  blue and orange square symbols correspond to the left and right
  panels of Fig.~\ref{advection2}, respectively. We separate
  identifiably distinct regions of behavior with dashed lines and
  number them with roman numerals.}
\label{variableP}
\end{figure}

\section{Including Tidal Dissipation: Paths to Misalignment}
\label{tidessec}

We now include tidal dissipation in the planet and allow the
semi-major axis of the planet to decay, and examine how the
non-dissipative regimes discussed in the previous section map onto
the final spin-orbit misalignment angle distributions. We use the standard
weak friction model for tidal dissipation; see SAL14 for details.

As the planet's semi-major axis decreases due to tidal dissipation,
the stellar spin precession rate increases (Eq.~\ref{alpha0}).  The LK
precession rate also increases, but not as dramatically, leading to an
overall gradual increase in both $\abar$ and $N_\max$.  In addition,
as the semi-major axis decays, the shape of the LK orbit changes, with
the minimum eccentricity of the LK cycles slowly increasing (see
Section 3.1 of ASL16); thus, the shape functions $\beta(\tau)$,
$\gamma(\tau)$ and $\psi(\tau)$ that drive the stellar spin dynamics
also slowly change in time. All of these changes, however, are slow
enough for the system to be treated as ``quasi-static'': For any given
LK cycle the spin dynamics of the initially-aligned trajectory is
still governed by the non-dissipative Hamiltonian, as described in the
previous section. However, over the orbital decay time (involving many
LK cycles), the coefficients
($N_\max$, $\abar$, $\bbar$, $\beta_M$,$\gamma_M$,$\psi_M$) in the Hamiltonian slowly change, and the
background non-dissipative phase space slowly evolves through
a sequence similar to that depicted in Figs.~\ref{samplephasespaces},
\ref{influence} and \ref{influence87}.

There are two consequences of this slow evolution of the Hamiltonian
coefficients. First, since at any given time the system is still
governed by a Hamiltonian, a (non-chaotic) trajectory still cannot
cross a separatrix (unless it has no choice -- more on this
later). The implication is that if a trajectory starts out
inside a certain separatrix, its behavior will continue to be governed
by that (slowly evolving) separatrix.  Second, because the system
evolves so gradually (or {\it adiabatically} - not to be confused with
the adiabatic regime!), an adiabatic invariant emerges: the area
enclosed by a trajectory in the $\cos\thetaSL - (\phiSL-N\tau)$ space
is an approximately conserved quantity. Together, these two ideas
(avoidance of separatrix crossings, and conservation of area) are all
that is necessary to understand the dissipative system. Thus, in
principle, knowing what separatrix governs the behavior of the
initially-aligned system at $t=0$, plus the knowledge of how that
separatrix changes under the influence of tidal evolution, should be
enough to determine the fate of the trajectory.

The above considerations indicate that the governing separatrix at
$t=0$ determines the fate of an initially aligned system.  The initial
separatrix is specified by the parameters (see Section 4.3)
\be
\abar_0 \equiv \abar (t=0), \quad N_{\rm max,0} \equiv N_\max (t=0).
\ee
In the presence of tidal dissipation, we expect that all possible
outcomes of the system may be classified using these two parameters.

\subsection{Varying the stellar spin period}

We begin by repeating the experiment of Section \ref{nondiss},
including the influence of tides
and numerically integrating the system until the planet orbit decays and circularizes. 
We select two initial
($t=0$) ``shapes'' for the LK orbit by setting $M_p=5 M_J$, $a_0=1.5$
AU (the initial semi-major axis), $a_b=300$ AU, and either
$\thetaLBo=89^\circ$ or $\thetaLBo=87^\circ$. For $t > 0$, the
``shape'' of the orbit slowly evolves (as the semi-major axis
shrinks). As in Section \ref{nondiss}, we vary the stellar spin
period (which remains constant throughout the tidal evolution). We set
$\theta_{\rm sl,0} = 0$ and plot $\theta_{\rm sl,f}$, the final
spin-orbit misalignment angle, as a function of the stellar spin
period in Figure \ref{variableP}. As before, we cover a somewhat
unphysical range of stellar spin periods in order to capture all
possible behaviors and outcomes.

Five distinct categories of outcomes in Fig.~\ref{variableP} can be
identified:

({\bf I}) The distinct bimodal distribution at very high
spin periods ($P_\star \go 23$ days and $P_\star \go 9$ days in
Fig.~\ref{variableP} upper and lower panels, respectively);

({\bf II}) The unimodal, monotonically decreasing distribution at high to
intermediate spin periods ($7 \lo P_\star \lo 20$ days in
Fig.~\ref{variableP}, upper panel);

({\bf III}) The region of wide-spread chaos ($0.5 \lo P_\star \lo 9$ days in 
Fig.~\ref{variableP} lower panel);

({\bf IV}) The diagonally striated pattern of spin-orbit
misalignments at low and very low spin periods ($0.5 \lo P_\star \lo
7$ days in Fig.~\ref{variableP} upper panel); we term this behavior
``adiabatic advection'' and discuss it extensively below;

({\bf V}) The region of very exceptionally small final misalignments at very low spin
periods ($P_\star \lo 0.4$ days and $P_\star \lo 0.3$ days in the
upper and lower panels of Fig.~\ref{variableP}, respectively). 

\begin{figure*}
\includegraphics[width=\textwidth]{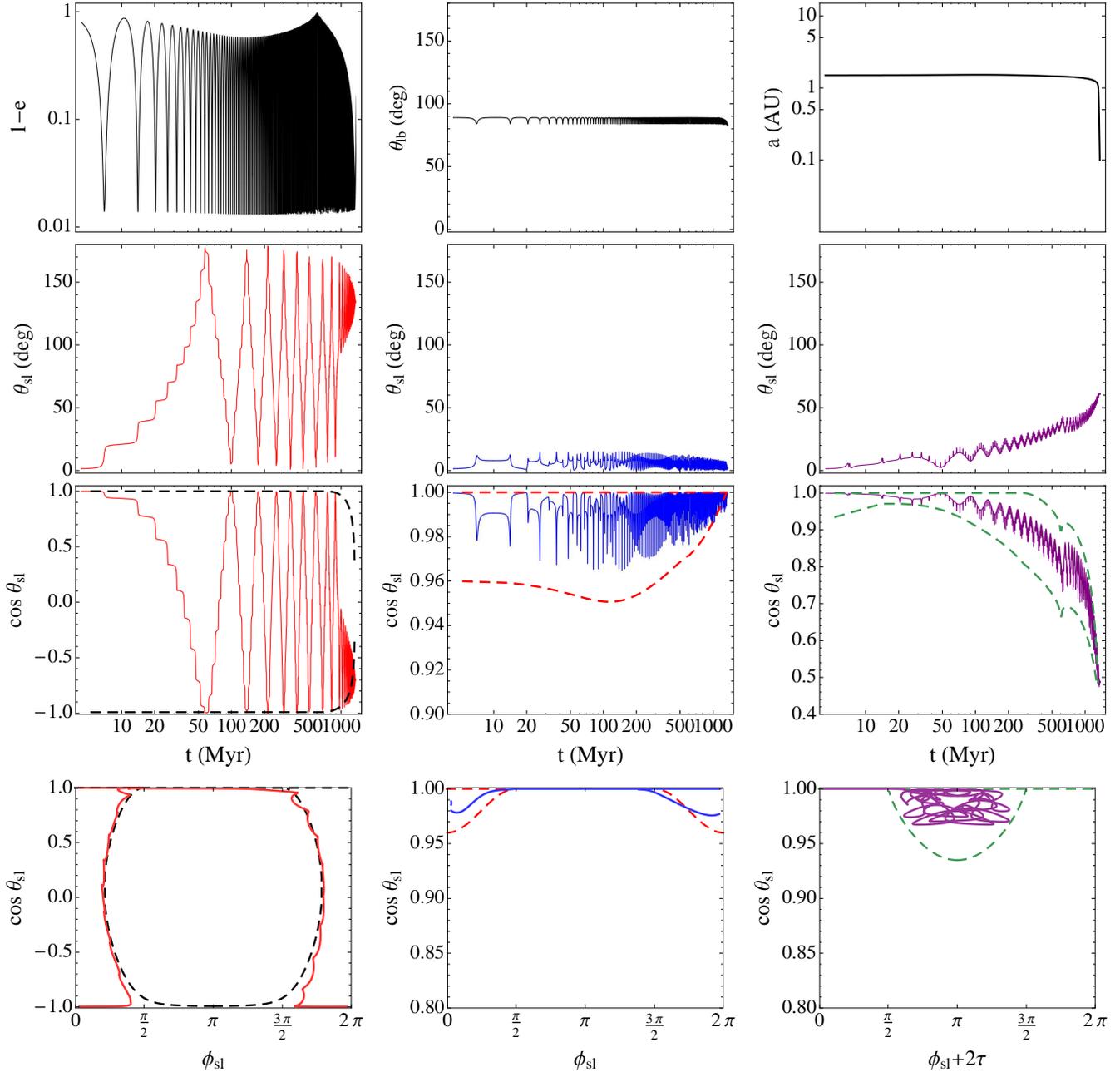}
\caption{Sample evolution trajectories, including tidal dissipation,
  for $\thetaLBo=89^\circ$, $M_p=5M_J$ and three different values of
  the stellar spin period. Left panels: $P_\star=30$ days; middle
  panels: $P_\star=7.07$ days; right panels: $P_\star=1.67$ days. Top
  row: evolution of the orbital elements (the planet's eccentricity,
  orbital inclination relative to the outer binary, and the semi-major
  axis). The orbital elements' evolution is independent of the stellar
  spin period and is therefore the same for all three cases. Second
  row: evolution of the spin-orbit misalignment angle
  $\thetaSL$. Third row: evolution of $\cos\thetaSL$ (solid lines) as
  well as the relevant background separatrix (dashed lines; see next). 
Bottom row: the initial background phase space separatrices (dashed
lines; see Fig \ref{samplephasespaces}) as well as the first full
cycle of evolution of the trajectory (solid lines), showing that the
relevant background separatrices, from left to right, are the $N=0$
center island (shown in black-dashed line), the $N=0$ top island
(shown in red-dashed line), and the $N=-2$ resonance (shown in
green-dashed line).}
\label{samplepaths}
\end{figure*}

Figure \ref{samplepaths} illustrates the time evolution for the
behaviors ({\bf I}), ({\bf II}) and ({\bf IV}).  We discuss each
category of outcomes individually in the following subsections.

\subsection{Non-adiabatic behavior: bimodality (I)}

We first address the clean bimodal spin-orbit misalignment
distribution found at high stellar spin periods in
Fig.~\ref{variableP} (upper panel). In order to understand its origin,
we need to know which separatrix governs the behavior of the
$\theta_{\rm sl,0} = 0$ trajectory at $t=0$. Two simple clues point 
to the answer: First, the left panels of Fig.~\ref{samplepaths} show
an example of the time evolution in the bimodal regime, as
well as the $t=0$ phase space for that evolution. We see that
the separatrix governing the behavior of the trajectory at
$t=0$ is the $N=0$ central island. Second, in Fig.~\ref{variableP} the
bimodal region ends very close to $\abar_0 \simeq 1$, i.e. at the
stellar spin period for which at $t=0$ we have $\abar \simeq 1$. From
Section \ref{nondiss}, we know that $\abar \simeq 1$ corresponds to
the transition between the non-adiabatic and adiabatic behavior, and that
at $\abar \simeq 1$ the governing separatrix for the initially aligned
system switches from being the $N=0$ center island to the $N=0$
top island (see Fig.~\ref{samplephasespaces} and
Fig.~\ref{influence}). Thus, the bimodality found at
high stellar spin periods in Fig.~\ref{variableP} corresponds to the
initially non-adiabatic regime.

To understand why the outcome is bimodal, we now need to know two
things: how the $N=0$ center separatrix changes in time (due to
tidal dissipation), and how the trajectory interacts with it.

The $N=0$ center separatrix evolves in time (Fig.~\ref{samplepaths},
left, the third panel from the top) in a way exactly analogous to the
sequence shown in Figs.~\ref{samplephasespaces} and \ref{influence}:
Initially the separatrix is attached to $\cos\thetaSL=1$. As tidal
dissipation acts to reduce the semi-major axis and increase $\abar$
and $N_\max$, the separatrix detaches from $\cos\thetaSL=1$ and slowly
shrinks.
On the other hand, the actual trajectory cannot shrink, due to the
aforementioned adiabatic invariance of the area it encloses in the
phase space.  Its initial area is set by the area of the separatrix at
$t=0$. At intermediate times the separatrix actually expands
(analogously to the transition between panels {\bf a} and {\bf b} of
Fig.~\ref{samplephasespaces}) and the trajectory remains inside the
separatrix. After the separatrix detaches and begins to shrink again,
there comes a point when the area of the separatrix is equal to the
area of the trajectory. At that point, the trajectory has no choice
but to cross the separatrix.

\begin{figure*}
\includegraphics[width=\textwidth]{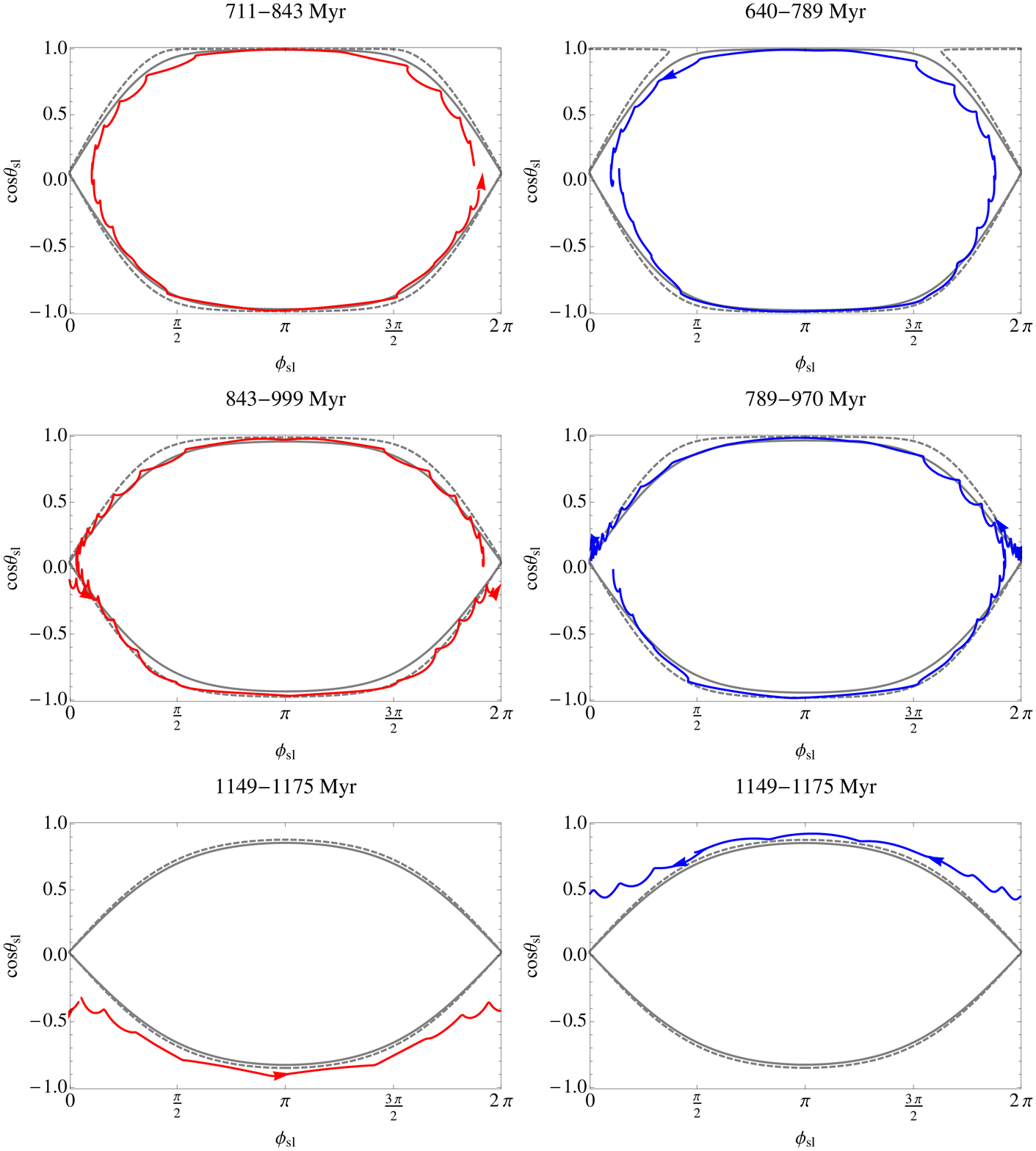}
\caption{Demonstration of the process that gives rise to the
  bimodality found in the non-adiabatic regime of Fig.~\ref{variableP}
  (top panel).  In each panel, the dashed (solid) grey line shows the
  $N=0$ separatrix at the beginning (end) of the time interval
  indicated at the top of the panel, while the colored line shows the
  actual time evolution of $\cos\thetaSL$ during that time
  interval. The red line (left panels) has $P_\star=30$ days and is
  the trajectory shown in the left panels of Fig.~\ref{samplepaths}
  and marked with a red X in Fig.~\ref{variableP}. The blue line
  (right panels) has $P_\star=29.87$ days and is marked with a blue O
  in Fig.~\ref{variableP}. Top panels: both trajectories are contained
  within the $N=0$ separatrix and their areas are smaller than the
  separatrix area. Middle panels: the $N=0$ separatrix has shrunk such
  that its area now matches the areas of the trajectories, thus the
  trajectories have no choice but to exit the resonance. The red
  (blue) trajectory's location at time of exit is such that it exits
  below (above) the separatix. Bottom panels: both trajectories are
  now caught in their respective part of phase space.}
\label{bimodality}
\end{figure*}

Figure \ref{bimodality} illustrates this idea. At time of crossing,
the trajectory can cross either the top or the bottom part of the
separatrix, depending on its phase. Two trajectories that start very
close together can, over time, accumulate enough difference in phase
that one ends up exiting through the top, and the other through the
bottom. This is the origin of the bimodality seen in
Fig.~\ref{variableP}.

This ``bifurcation'' phenomenon is analogous to the case of a pendulum
whose length is slowly decreased with time. The shorter the pendulum
gets, the larger its amplitude of oscillation becomes, until at some
point it must transition to circulating rather than oscillating. At
that point, the pendulum will ``choose'' to circulate either clockwise
or counterclockwise -- corresponding to a positive or negative
conjugate momentum -- depending on its phase at the time of
transition.

Finally, we note that while the bimodality is very clear and distinct
in the upper panel of Fig.~\ref{variableP}, it becomes increasingly
disordered at lower periods in the lower panel of
Fig.~\ref{variableP}. This is the onset of widespread chaos that
arises from resonance overlaps, as discussed in Section \ref{IATsub}
(see SL15).

\subsection{Stationary adiabatic behavior (II)}

For $\abar_0 \go 1$ and $N_{\rm max,0} \lo 0.5$, Fig. \ref{variableP}
(top panel) shows a smooth unimodal distribution of final spin-orbit
misalignments, with $\theta_{\rm sl,f}$ decreasing with decreasing
$P_\star$. In Section 5.2, we have already determined that
the behavior in this regime must be governed by the $N=0$ top side
island. The middle bottom panel of Fig.~\ref{samplepaths} shows that
this is indeed the case. In order to understand the distribution of
the final misalignments, we again need to ask how the $N=0$ top separatrix
evolves with time, and how the trajectory interacts with it.

Based on Fig.~\ref{influence}, we know that an increase in $\abar$
or $N_\max$ leads to a rapid decrease in the width of the $N=0$ top
island. On the other hand, again, the actual trajectory area is constant
and set by the initial area of the top island. Thus, as soon as any
significant semi-major axis decay occurs, the trajectory area will 
exceed the area of the top island, and the actual trajectory should
circulate on the outside of the island.
For a circulating trajectory, the new conserved area becomes the area
between the trajectory and the $\cos\thetaSL=0$ axis. Thus, as more
semi-major axis decay occurs, the trajectory cannot move up/down, it
can only straighten out, eventually settling on a constant
$\cos\theta_{\rm sl,f}$ equal approximately to its mean $\cos\thetaSL$
value at the time of decoupling from the top island.

Thus, if the top island is initially extended, the final $\theta_{\rm sl,f}$ 
should be (relatively) large. As the stellar spin period
decreases, the top island decreases in size; therefore, 
$\theta_{\rm sl,f}$ gets closer and closer to $0$.

A caveat to the above discussion is the following: Further into the
adiabatic regime, the assumption that only the $N=0$ Hamiltonian
determines the spin dynamics becomes increasingly erroneous, since the
stellar spin vector now precesses at at rate comparable to the
precession of $\hatL$ and thus sees more than just the time average of
the forcing functions. Thus, although the above discussion is
suggestive and sound, the actual behavior of the trajectory can be
more random and need not obey these rules. This is why, for example,
the trajectory depicted in the middle
column of Fig.~\ref{samplepaths} actually remains level with, or
inside, the top island for most of its evolution.

We term this regime of behavior ``stationary adiabatic'' because the
trajectory essentially cannot move away from its initial
location. This is to be contrasted with type ({\bf IV}) behavior
discussed below.

Note that type ({\bf II}) behavior need not always be present: for example, there is no such apparent behavior in the bottom panel of Fig. \ref{variableP}. This is because, in many cases, it can be replaced by wide-spread chaos, as discussed below.

\subsection{Widespread Chaos (III)}

For intermediate values of the stellar spin period, in the lower panel
of Fig.~\ref{variableP} we can identify a region of widespread or near
widespread in the final spin-orbit misalignments, as a result of
chaotic spin evolution. This type of behavior was extensively studied
in SL15 and we give only a short discussion here.

As discussed in SL15, chaotic behavior often arises from resonance
overlaps. In Section \ref{IATsub}, based on Fig.~\ref{influence87}
which shows the existence of resonance overlaps for certain spin
periods, we suggested that misalignment distributions computed using
this particular set of ``shape'' functions should exhibit chaotic
regions. This is indeed the case. In fact, there is good agreement
between Fig.~\ref{influence87} and the lower panel of
Fig.~\ref{variableP}: For example, for $3.5 \lo P_\star \lo 5$ days
the behavior of the trajectory should be governed by the $N=-1$
resonance and
this resonance does not overlap with the $N=0$ island. Thus, there is
no chaos in this period range.

In general, we expect the appearance of chaos to be correlated with
the widths of the resonances. Since the widths of the resonances
generally decrease with increasing $\abar_0$ (see SL15), we expect the 
chaotic bands to be confined to lower values of $\abar_0$.

We note, however, that we cannot readily explain the existence of
widespread chaos in the $P_\star > 5$ days region of
Fig.~\ref{variableP}. This is because our resonance overlap analysis
(SL15) assumes that the system is already quite adiabatic, so that any
non-adiabatic effects can be treated perturbatively. For $P_\star > 5$
days this is not the case, since $\abar_0$ is close to or 
less than $1$.

\subsection{Adiabatic Advection (IV)}

\begin{figure}
\centering
\scalebox{0.67}{\includegraphics{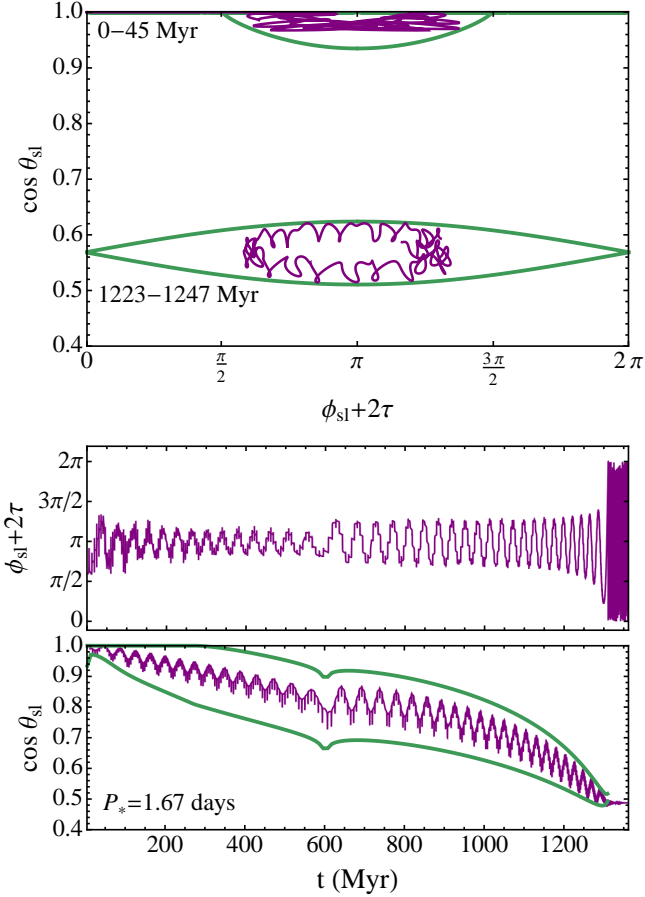}}
\caption{Demonstration of the process of adiabatic advection by an
  $N=-2$ resonance. In both panels, the purple lines show the actual
  time evolution of the spin trajectory, corresponding to the right
  panels of Fig.~\ref{samplepaths}, and is marked with a purple X in
  Fig.~\ref{variableP}. Top panel: the median shape of the $N=-2$
  resonance and the time evolution of $\cos\thetaSL$ is shown at two
  different time intervals. In both intervals, the actual trajectory
  is contained inside the resonance. Bottom panel: the full evolution
  of $\phi_{\rm sl}+2\tau$ (top sub-panel) and $\cos\thetaSL$ (bottom
  sub-panel, purple) as well as the maximum width of the $N=-2$
  resonance vs time (bottom sub-panel, green), confirming that the
  trajectory advects with the resonance until the resonance shrinks
  significantly and the trajectory has no choice but to exit.}
\label{advection}
\end{figure}

The concept of adiabatic advection was first considered by SL15 in a schematic manner.
Here we demonstrate that 
adiabatic advection indeed applies in the realistic situation of ``LK oscillation
+ tidal decay'' considered in this paper. Adiabatic advection
is a novel way of generating spin-orbit misalignment.

The idea of adiabatic advection is simple: If at $t=0$ the behavior of
the $\cos\theta_{\rm sl,0}=1$ trajectory is governed by a resonance
with a specific $N$ ($=-1,-2,\cdots$), then as tidal dissipation
reduces the semi-major axis of the orbit, the trajectory can be
advected by the governing resonance to non-zero misalignments. Figure
\ref{advection} shows an example of this behavior: in the top panel,
at $t=0$ we see that the trajectory is trapped inside the $N=-2$
resonance, which is still ``attached'' to $\cos\thetaSL=1$. Over a Gyr
later, the resonance has detached and moved down significantly, and
the trajectory has likewise moved down and remains inside the
resonance, producing a significant spin-orbit misalignment.

\begin{figure}
\centering
\scalebox{0.67}{\includegraphics{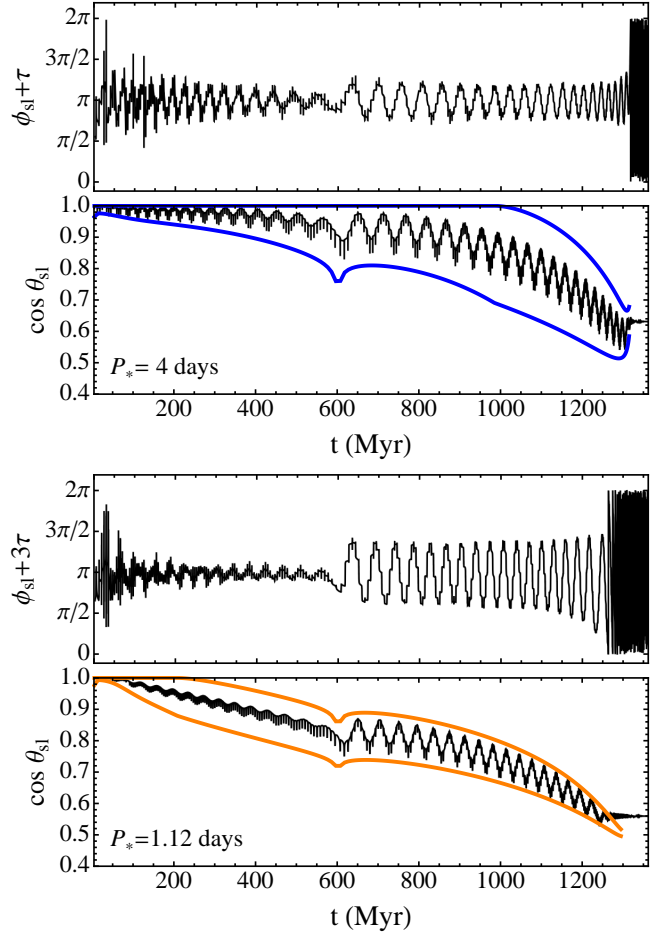}}
\caption{Demonstration of the process of adiabatic advection by the
  $N=-1$ (top) and the $N=-3$ (bottom) resonance. The top (bottom)
  time evolution is marked in Fig.~\ref{variableP} with a blue
  (orange) square symbol. As in Fig. \ref{advection} the two sample
  trajectories remain trapped in their respective resonances (top:
  $N=-1$, shown in blue; bottom: $N=-3$, shown in orange) and are
  advected with them.}
\label{advection2}
\end{figure}

\begin{figure*}
\includegraphics[width=\textwidth]{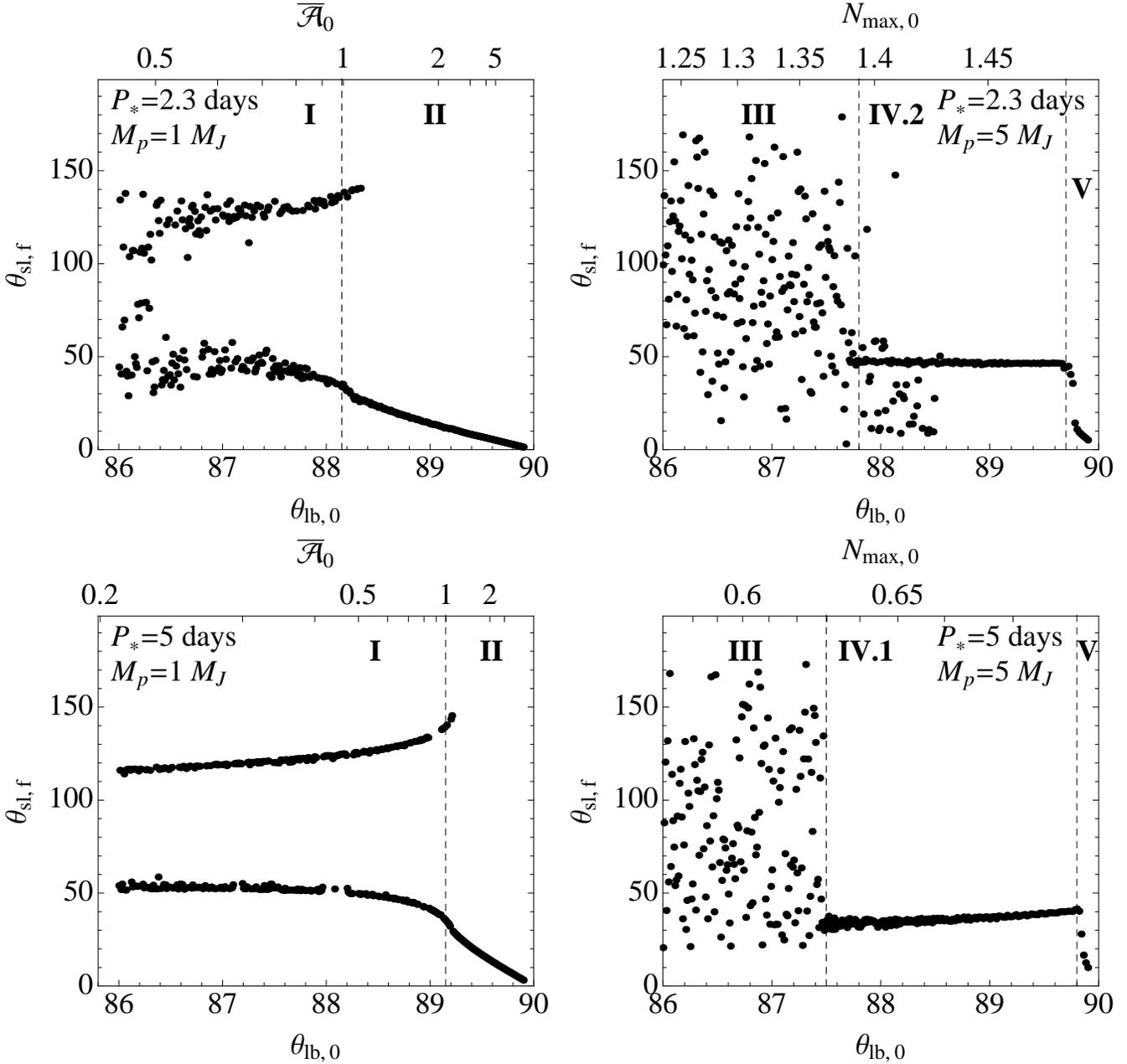}
\caption{Final spin-orbit misalignment angle as a function of initial
  orbital inclination $\thetaLBo$, for two different planet masses and
  two stellar spin periods. The $M_p=1 M_J$ panels (left) exhibit
  non-adiabatic bimodal ({\bf I}) and stationary adiabatic ({\bf II})
  behaviors, with the transition from one to the other occuring, as
  expected, at $\bar{\mathcal{A}}_0\simeq 1$. The $M_p=5 M_J$ panels
  (right) exhibit chaos ({\bf III}) and adiabatic advection ({\bf IV})
  by the $N=-2$ (top) and $N=-1$ (bottom) resonances. Because $N_{\rm
    max,0}$ is rather insensitive to the initial inclination (see
  Fig.~\ref{Nmaxvsi}), the final misalignment angle after advection is
  nearly independent of $\thetaLBo$. The higher mass planets also show
  a full adiabatic region ({\bf V}).}
\label{thetaslfvsi}
\end{figure*}

As another means of looking at the situation, we note that at the
center of the $N$-th resonance we have, by definition,
$\phiSL-N\tau=0$. Thus, a trajectory trapped inside the resonance
librates about this point; that is, $\phiSL-N\tau$ of such a
trajectory should exhibit moderate variation about $0$. In the lower
panel of Fig.~\ref{advection}, we show that this is indeed the case:
$\phiSL+2\tau$ oscillates about $0$, indicating that the trajectory is
trapped inside the $N=-2$ resonance. Likewise, the oscillations of
$\cos\thetaSL$ never exceed the maximum width of the $N=-2$ resonance,
demonstrating that the trajectory is always contained inside.

Similarly, Figure \ref{advection2} shows sample advections by the
$N=-1$ (top) and $N=-3$ (bottom) resonances. By locating these
examples on the $\theta_{\rm sl,f}$ vs $P_\star$ plot
(Fig.~\ref{variableP}, upper panel), we conclude that each of the
diagonally striated lines in the top panel of Fig.~\ref{variableP},
located at $0.5 \lo N_{\rm max,0} \lo 1$, $1 \lo N_{\rm max,0} \lo 2$,
etc., corresponds to advection by a resonance of a different $N$.

\subsection{Fully Adiabatic Evolution (V)}

For very low (extremely fast) spin periods, the systems presented in
both panels of Fig.~\ref{variableP} are fully adiabatic. 
(We note that in the particular 
case of Fig.~\ref{variableP} the spin periods in this regime are unphysically 
short; however, the same behavior can be achieved at more reasonable spin periods
by changing the system orbital parameters or planet mass.) 
In this
regime, the stellar spin vector has no trouble following the planet
angular momentum vector as it undergoes LK oscillations, and no
spin-orbit misalignment is generated. We note that, while we
understand what happens in this regime, it is difficult to pinpoint
its onset, i.e. the value of $P_\star$ or perhaps $N_\max$ for which
it starts.

\section{Predictive Power of the Theory}

\begin{figure}
\centering
\scalebox{0.67}{\includegraphics{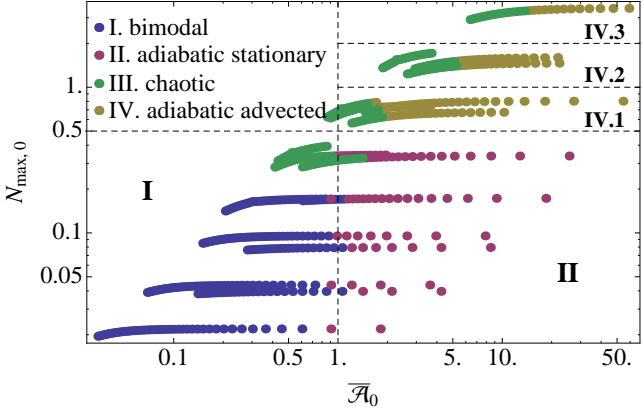}}
\caption{Compilation of the outcomes of a suite of time evolutions
  with variable initial conditions and parameters, plotted in the
  $N_{\rm max,0}$ vs $\bar{\mathcal{A}}_0$ space. Each outcome is
  classified into the categories discussed in Section
  \ref{tidessec}. With dashed lines and roman numerals we denote which
  type of outcome we {\it expect} in that region of parameter
  space. The variable initial conditions/parameters include
  combinations of various planet masses, stellar spin periods, binary
  separations, and initial orbital inclinations. Despite varying all
  of these parameters, we see that the outcomes still agree with our
  expectations based on Section \ref{tidessec}: The non-adiabatic
  bimodal ({\bf I}) and adiabatic stationary ({\bf II}) behaviors
  occur at low $N_{\rm max,0}$, and are separated by
  $\bar{\mathcal{A}}_0\simeq 1$. At higher $N_{\rm max,0}$, adiabatic
  advection ({\bf IV}) and chaotic behavior ({\bf III}) become
  possible, with the chaotic behavior being restricted to lower values
  of $\bar{\mathcal{A}}_0$ (for which, generally speaking, all the
  resonances are wider and more likely to overlap).}
\label{params}
\end{figure}

In the previous two sections, we have focused on numerical experiments
that change the spin dynamics in the simplest way possible: by
varying the stellar spin period, while keeping all other system
parameters fixed. We would now like to check whether the understanding
of the different outcomes for $\theta_{\rm sl,f}$ we have developed in the
previous sections holds up when we vary some other parameters of the system.
Thus, in Fig.~\ref{thetaslfvsi} we present the
distributions of final spin-orbit misalignment angles as a function of
the initial orbital inclination $\thetaLBo$, for two values of the
stellar spin period and two planet masses.

We find that, indeed, our classification of the misalignment outcomes 
based on $\abar_0$ and $N_{\rm max,0}$ holds up
well. For Jupiter-mass planets, the outcomes are either bimodal (type
{\bf I}) or stationary adiabatic (type {\bf II}), with the transition
between the two regimes occuring at $\abar_0 \simeq 1$, as
expected. For the heavier, $5 M_J$ planets, a chaotic band ({\bf III})
appears at lower inclinations, likely due to resonance overlaps. Aside
from that, however, in the non-chaotic regions we find that
$\theta_{\rm sl,f}$ is nearly constant, consistent with the fact that
$N_{\rm max,0}$ is moderately large and nearly independent of
$\thetaLBo$ (as expected based on Fig. \ref{Nmaxvsi}). 
Based on the values of $N_{\rm max,0}$, we infer 
that the top right panel of Fig.~\ref{thetaslfvsi} shows an extended
region of adiabatic advection by the $N=-2$ resonance (type {\bf IV.2}
behavior), whereas the bottom right panel of Fig.~\ref{thetaslfvsi}
shows a region of advection by the $N=-1$ resonance (type {\bf IV.1}
behavior).

Finally, taking one step further, we
carry out a suite of calculations for the ``LK oscillation + tidal
decay'' time evolutions, with various initial conditions and
parameters (different planet masses, stellar spin periods, binary
separations, and initial orbital inclinations). We categorize each
time evolution according to its misalignment outcome: bimodal ({\bf
  I}), stationary adiabatic ({\bf II}), chaotic ({\bf III}) or
adiabatically advected ({\bf IV}), and plot these outcomes in the
$N_{\rm max,0}$ vs $\abar_0$ space (Fig. \ref{params}). From this
figure, it is clear that our understanding of the different outcomes
is sound: at $\abar_0 \lo 1$, bimodality dominates; at $\abar_0 \go 1$
but $N_{\rm max,0} \lo 0.5$ stationary adiabatic behavior is
prevalent; for $\abar_0 \go 1$ and $N_{\rm max,0} \go 0.5$, adiabatic
advection is dominant, except where the evolution is chaotic, with the
chaotic evolution restricted to lower values of $\abar_0$.

We conclude that in general,
the two parameters, $\abar_0$ and $N_{\rm max,0}$, completely determine
the evolutionary behavior of the spin-orbit misalignment and
whether the final misalignment angle of a given system can be
high. For $\abar_0 \lo 1$ (regime {\bf I}), the final misalignment
distribution is bimodal, and thus a system is equally likely to have
low misalignment as high misalignment. For $\abar_0 \go 1$ but $N_{\rm max,0} 
\lo 0.5$ (regime {\bf II}) the system is incapable of
achieving a significantly misaligned state. For $\abar_0 \go 1$ and
$N_{\rm max,0} \go 0.5$ a calculation of resonance widths must be
carried out to determine whether the system is chaotic ({\bf III}) or
advecting ({\bf IV}), but in general it can be expected that if
$\abar_0$ is relatively high then the system is not chaotic, and will
attain non-zero, but strictly prograde and fairly modest,
misalignment.

\section{Discussion}

\subsection{Complications due to Spin Feedback on Orbit}

As explained in Section \ref{SRFsec}, a major assumption in the
analysis laid out in this paper, as well as in SL15, is the omission
of the extra precession the planet's orbit experiences due to
perturbation from the stellar quadrupole.  This omission enables us to
considerably simplify the spin dynamics problem by reducing it to a 1D
Hamiltonian system.
Stellar feedback on planet's orbit becomes important if the host star
has nearly as much, or more, angular momentum as the planet's orbit
(see Eq.~\ref{eq:SoverL}).  Thus, one may question whether our
analysis is truly applicable to Jupiter-mass (as opposed to heavier)
planets and rapidly rotating stars. However, in ASL16 we have run a
comprehensive suite of numerical simulations including stellar
feedback on the orbit and all other important effects.  We found that,
while under certain conditions the bimodal and stationary adiabatic
behaviors expected to dominate for Jupiter-mass planets can be
disrupted, on the whole, bimodality remains nearly ubiquitous.  We
conclude that our classification of the different misalignment
evolution modes and outcomes is generally valid.

\subsection{Stellar Spindown}
In this work, we have assumed that the stellar spin remains constant 
throughout the proto-HJ's tidal decay and circularization. In reality,
solar-type stars experience significant spindown due to magnetic winds.
This spindown acts to temporarily reduce the strength of the coupling 
between the stellar spin and the planet orbit, decreasing both $N_\max$
and $\bar{\mathcal{A}}$. However, it is still the initial values of 
these parameters ($N_{\rm max,0}$ and $\bar{\mathcal{A}}_0$) 
that set the qualitative behavior of the system. Some differences in the 
final value of the spin-orbit misalignment angle $\theta_{\rm sl,f}$ due
to stellar spindown could be expected, but the regime
classification and predictive power should remain unchanged.

\subsection{Primordial Misalignment}

In this paper we have focused on stellar spin dynamics in systems that
have no initial spin-orbit misalignment. However, since several ways
of generating primordial misalignment have been proposed (see
references in Section 1), the dynamics of initially misaligned systems
are of potential interest.

Although a thorough exploration of the dynamics of initially
misaligned systems is beyond the scope of this work, we believe the
ideas developed in this paper, particularly the importance of the parameters
$\abar_0$ and $N_{\rm max,0}$, are still applicable to initially
misaligned systems. The fate of such systems should still be determined
by the initial phase space of the system and where in that phase
space the system is initialized. For example, for trajectories
starting anywhere inside the $N=0$ center island
(see Fig.~\ref{samplephasespaces}), the final outcome should be
bimodal, just as for initially aligned systems; the only difference is that, for
an initially misaligned system, the 
peaks of the bimodal distribution of the final misalignment angle 
would lie closer to $90^\circ$ due to the smaller initial area of the trajectory. In 
Fig.~$26$ of ASL16 we have demonstrated that this is the case.

Thus, the frameworks developed in this paper for the special case of
initially aligned systems can be easily generalized to systems with
arbitrary initial misalignments and phases.

\section{Summary}

We have developed an analytical theory to explore and classify the
various regimes of stellar spin dynamics driven by planets undergoing
Lidov-Kozai migration.  In our previous work (Storch \& Lai 2015) we
analyzed only the idealized non-dissipative Lidov-Kozai system in the
adiabatic regime (when the spin precession frequency is higher than
the LK oscillation frequency) and succeeded in explaining the origin
of chaotic behavior of stellar spin. In this paper we have
significantly expanded our analysis to include the effects of various
short-range forces (e.g., apsidal precession of the planet's orbit due
to General Relativity and tidal bulge) and considered all possible
spin dynamical regimes. Most importantly, we have included tidal
dissipation in the planet, which allows us to examine the long-term
evolution of the spin-orbit misalignment angle as the planet migrates
in semi-major axis and becomes a hot Jupiter. The work presented here
provides a solid theoretical understanding for the various pathways
toward spin-orbit misalignments as revealed in our extensive numerical
simulations (Storch, Anderson \& Lai 2014; Anderson, Storch \& Lai 2016).

We find that, in general, the behavior of a ``stellar spin + planet +
binary'' system with a given set of initial conditions, planet mass,
and stellar spin rate is governed primarily by two parameters (Section
4): $N_\max$ (Eq.~24), which compares $\bar\alpha$, the average
maximum precession frequency of the stellar spin, with the LK
eccentricity oscillation frequency, and $\abar$ (Eq.~26), which
compares $\bar\alpha$ with the averaged rate of change of the planet's
orbital angular momentum vector $\hatL$.  For $\abar \lo 1$ (which
implies $N_\max \ll 1$), the spin dynamics is in the non-adiabatic
regime. The adiabatic regime ($\abar \go 1$) is divided into two
sub-regimes with $N_\max \lo 1$ and $N_\max \go 1$ (recall that $\abar
\go N_\max$). Each of these regimes and sub-regimes has a different phase-space
resonance structure (see Figs.~5-7), which governs the evolution of the system.

In the presence of tidal dissipation (Section 5), $N_\max$ and $\abar$
vary slowly with time, but the fate of the system is entirely
determined by the values of these parameters at $t=0$ (i.e. during the
first LK cycle). In general, five distinct spin-orbit evolutionary
behaviors and outcomes are possible (see Section 5.1), and Fig.~8,~13
and 14 summarize these various types of outcomes.  We find that when
$\abar_0 \lo 1$ (implying $N_{\rm max,0} \ll 1$), the final spin-orbit
misalignment distribution is bimodal (type {\bf I}), and thus the
system is equally likely to settle into a retrograde or prograde orbit. 
When $\abar_0 \go 1$ and $N_{\rm max,0} \lo 0.5$, the
system experiences ``stationary adiabatic'' behavior (type {\bf II})
and cannot achieve very high (retrograde) misalignments. When When $\abar_0 \go 1$ and
$N_{\rm max,0} \go 0.5$, the system is either chaotic (type {\bf III})
or experiences ``adiabatic advection'' (type {\bf IV}), wherein it can
slowly accumulate a modest amount of spin-orbit misalignment (never
more than $90^\circ$). The chaotic regime is typically restricted to
lower values of $\abar_0$. At very high values of both $\abar_0$ and
$N_\max$ the system is fully adiabatic (type {\bf V}) and cannot
accumulate any misalignment.

Overall, the theoretical work presented in this paper and in Storch \&
Lai (2015) complements our own numerical Monte-Carlo studies of hot
Jupiter formation via Lidov-Kozai migration in stellar binaries
(Storch et al.~2014; Anderson et al.~2016) and those by others
(e.g. Fabrycky \& Tremaine 2007; Correia et al.~2011; Petrovich 2015).
Our theory provides a framework to understand some of the intriguing
numerical results on the final spin-orbit misalignment distributions.
For example, we have shown that the bimodal distribution, seen in some
of the simulations of Fabrycky \& Tremaine (2007), Correira et
al.~(2011) and Storch et al.~(2014), arises naturally from a
``bifurcation'' phenomenon associated with the crossing of a separatrix
in the phase space (see Section 5.2). Another example is the novel
``resonance advection'' phenomenon (see Section 5.5), which leads to
the production of misaligned systems even in ``adiabatic'' systems.
Finally, we note that, while we have focused on Lidov-Kozai migration
induced by a stellar binary in this paper, the concepts and methods
developed in this paper can also be applied to other high-eccentricity
migration scenarios involving planet-planet secular interactions 
(e.g. Nagasawa et al. 2008; Wu \& Lithwick 2011; Beaug{\'e} \& Nesvorn{\'y} 2012; 
Petrovich \& Tremaine 2016; Hamers et al. 2016).

\section*{Acknowledgments}
This work has been supported in part by NSF grant AST-1211061, and
NASA grants NNX14AG94G and NNX14AP31G. KRA is supported by the NSF Graduate 
Research Fellowship Program under Grant No. DGE-1144153. 
NIS acknowledges partial support through a Sherman Fairchild Fellowship at Caltech.

\end{document}